\documentclass[12pt]{article}
\pdfoutput=1
\usepackage[utf8]{inputenc}
\usepackage{amsfonts,amsmath}
\usepackage{jheppub}

\usepackage{color}
\usepackage{tabularx}

\usepackage{ifthen}
\usepackage{enumerate}
\usepackage{amsmath}
\usepackage{amssymb}
\usepackage[mathscr]{eucal}
\usepackage[errorshow]{tracefnt}
\usepackage{amsmath}
\usepackage{slashed}
\usepackage{amssymb}
\usepackage{array}
\usepackage{amsthm}
\usepackage{eucal}
\usepackage{epsf}
\usepackage{epsfig}
\usepackage{psfrag}
\usepackage{multirow}
\usepackage{wasysym} 
\usepackage{tikz-cd} 
\usepgfmodule{shapes}
\usetikzlibrary{backgrounds, arrows,calc,shapes,decorations.pathreplacing, decorations.markings, automata,positioning}

\newcommand{\nn}{\nonumber}
\newcommand{\rf}{r_{\phi}}

\newcommand{\beqa}{\begin{eqnarray}}
\newcommand{\eeqa}{\end{eqnarray}}

\def\W{{\cal W}} 
\def\Tr{\rm Tr}
\newcommand{\tr}{\text{tr}\,}

\newcommand{\beq}{\begin{equation}}
\newcommand{\eeq}{\end{equation}}
\newcommand{\bea}{\begin{eqnarray}}
\newcommand{\eea}{\end{eqnarray}}

\newcommand{\CB}{{\mathcal B}}
\newcommand{\CC}{{\mathcal C}}

\newcommand{\CF}{{\mathcal F}}

\newcommand{\CM}{{\mathcal M}}
\newcommand{\M}{{\mathfrak M}}
\newcommand{\CN}{{\mathcal N}}
\newcommand{\CO}{{\mathcal O}}

\newcommand{\CT}{{\mathcal T}}

\newcommand{\CW}{{\mathcal W}}
\newcommand{\CZ}{{\mathcal Z}}

\newcommand\qt{\tilde q}
\newcommand\bt{\tilde b}
\newcommand\pt{\tilde p}
\newcommand\Qt{\tilde Q}
\newcommand\Pt{\tilde P}

\newcommand\s{s_1}

\newcommand{\be}{\begin{equation}}
\newcommand{\ee}{\end{equation}}

\newcommand{\bpic}{\begin{tikzpicture}}
\newcommand{\epic}{\end{tikzpicture}}

\def\+{{+\!\!\!+}}

\def\a{\alpha} 
\def\b{\beta}

\def\0{\nonumber}

\def\W{{\cal W}}

\def\Tr{{\rm Tr}}

\def\1{{\bf 1}}

\title{Lagrangians for generalized Argyres-Douglas theories}

\preprint{SISSA 32/2017/MATE-FISI}

\author[1,2]{Sergio Benvenuti}
\author[2,3]{Simone Giacomelli}

\affiliation[1]{International School of Advanced Studies (SISSA), Via Bonomea 265, 34136 Trieste, Italy}
\affiliation[2]{INFN, Sezione di Trieste, Via Valerio 2, 34127 Trieste, Italy}
\affiliation[3]{International Center for Theoretical Physics, Strada Costiera 11, 34151 Trieste, Italy}
\emailAdd{benve79@gmail.com, sgiacome@ictp.it}

\abstract{We continue the study of Lagrangian descriptions of $\CN=2$ Argyres-Douglas theories. We use our recent interpretation in terms of sequential confinement to guess the Lagrangians of all the Argyres-Douglas models with Abelian three dimensional mirror. We find classes of four dimensional $\CN=1$ quivers that flow in the infrared to generalized Argyres-Douglas theories, such as the $(A_k,A_{kN+N-1})$ models. We study in detail how the $\CN=1$ chiral rings map to the Coulomb and Higgs Branches of the $\CN=2$ CFT's. The three dimensional mirror RG flows are shown to land on the $\CN=4$ complete graph quivers. We also compactify to three dimensions the gauge theory dual to $(A_1,D_4)$, and find the expected Abelianization duality with $\CN=4$ SQED with $3$ flavors.}

\begin{document}

\maketitle
\section{Introduction and summary}

We recently \cite{Benvenuti:2017lle,Benvenuti:2017kud} found physical interpretations of the $\CN=1$ Lagrangians for $(A_1,A_{2N-1})$ Argyres-Douglas (AD) theories discovered in \cite{Maruyoshi:2016tqk, Maruyoshi:2016aim} by Maruyoshi and Song. After appropriately correcting the Lagrangians, in order to account for unitarity violations and chiral ring stability, we obtained complete and consistent theories, which for instance allow to study the chiral rings and the moduli space of vacua, that indeed map to the Coulomb and Higgs branches of the $\CN=2$ AD CFT's.

The consistent description found in \cite{Benvenuti:2017lle,Benvenuti:2017kud} allows to compactify on a circle, and in $3d$ we discovered  two different physical interpretations. First, in $3d$ the $\CN=2$ theories are dual to an Abelian model with enhanced supersymmetry. Second, on the mirror side, using a duality for $3d$ $\CN=2$ theories with monopoles in the superpotential \cite{Benini:2017dud}, entire quiver tails  "sequentially confine" and the RG flow lands on $\CN=4$ SQED with $N$ flavors, which was indeed predicted to be the $3d$ mirror of $(A_1,A_{2N-1})$ \cite{Nanopoulos:2010bv,Xie:2012hs,Boalch}. 

In this paper we generalize the story to all the AD models with a known $3d$ Abelian mirror. Such theories can be obtained wrapping $k+1$ $M5$ branes  on a sphere with
\begin{itemize}
\item an irregular puncture (this class is called $(A_k,A_{kN+N-1})$ \cite{Cecotti:2010fi}),
\item an irregular puncture and a minimal puncture. 
\end{itemize}
Their $3d$ mirrors generalize $\CN=4$ SQED with $N$ flavors to "complete graphs" Abelian quivers with $k$ and $k+1$ gauge nodes, respectively \cite{Nanopoulos:2010bv,Xie:2012hs,Boalch} .

Our guiding principle to identify the Lagrangian descriptions for the above two classes of AD models is the ``sequential confinement'' $3d$ RG flow, which in the case of $(A_1,A_{2N-1})$ has been described in detail in \cite{Benvenuti:2017kud}. 

First we need to find $3d$ $\CN=4$ mirror pairs $\CT_{UV} \leftrightarrow \tilde{\CT}_{UV}$ with the property that in both $\CT_{UV}$ and $\tilde{\CT}_{UV}$ all non-Abelian gauge groups are balanced, i.e. they have $N_f=2N_c$. We find two classes of such mirror pairs: $\CT_{UV}$ is a linear quiver, while  $\tilde{\CT}_{UV}$ is a star-shaped quiver. The ``sequential confinement'' mirror RG flow starts from $\tilde{\CT}_{UV}$ and lands on the $\CN=4$ complete graphs  \cite{Nanopoulos:2010bv,Xie:2012hs,Boalch} dual to the above two classes of AD models.

 On the other hand, $\CT_{UV}$ can be uplifted to a $\CN=2$ superconformal linear quiver in $4d$, and provides the UV starting point for the Maruyoshi-Song flow to the $4d$ $\CN=1$ Lagrangians we are looking for.


The strategy outlined above allows us to find $\CN=1$ Lagrangians for all the AD theories with Abelian $3d$ mirror. For instance, our method leads to the prediction that the $4d$ Lagrangian description of $(A_k,A_{kN+N-1})$ is the following $\CN=1$ quiver with $k$ $SU(n_i)$ gauge groups
  \be\label{QUIVER} \bpic  \path  (-2.5,0) node[circle,draw](x1) {$\,N\,$} -- (-0.5,0) node[circle,draw](x2) {$2N$} -- (1.5,0) node(x3) {$\cdots$} -- (3.5,0)  node[circle,draw](x5) {$k N$} -- (5.5,0) node[rectangle,draw](x6) {$\,1\,$} ;
   \path  (-2.5,0) node[circle,draw] {$\,\quad$} -- (-0.5,0) node[circle,draw] {$\,\,\quad$} -- (3.5,0)  node[circle,draw] {$\,\,\quad$};
\draw [->] (x1) to[bend left] (x2); \draw [<-] (x1) to[bend right] (x2);
\draw [->] (x2) to[bend left] (x3); \draw [<-] (x2) to[bend right] (x3);
\draw [->] (x3) to[bend left] (x5); \draw [<-] (x3) to[bend right] (x5);
\draw [->] (x5) to[bend left] (x6); \draw [<-] (x5) to[bend right] (x6);
\draw [->] (x1) to[out=-45, in=0] (-2.5,-1) to[out=180,in=225] (x1);
\draw [->] (x2) to[out=-45, in=0] (-0.5,-1) to[out=180,in=225] (x2);
\draw [->] (x5) to[out=-45, in=0] (3.5,-1) to[out=180,in=225] (x5);
\node[below right] at (-2.4,-0.6){$\phi_1$};
\node[below right] at (-0.4,-0.6){$\phi_2$};
\node[below right] at (3.6,-0.6){$\phi_k$};
\node[above right] at (-1.8,0.3){$b_1$};
\node[above right] at (0.2,0.3){$b_2$};
\node[above right] at (2.0,0.3){$b_{k-1}$};
\node[below right] at (-1.7,-0.3){$\bt_1$};
\node[below right] at (0.3,-0.3){$\bt_2$};
\node[below right] at (2.0,-0.3){$\bt_{k-1}$};
\node[above right] at (4.2,0.3){$q$};
\node[below right] at (4.3,-0.3){$\qt$};
  \epic\ee 
Once the quiver is known, it is a trivial task to perform A-maximization and find $4d$ checks of the proposal. The consistent superpotential turns out to be
 \be\label{superpot} \CW_{4d}= \sum_{i=1}^k tr(\phi_i (b_i \bt_i-b_{i-1}\bt_{i-1}))  + \sum_{r=0}^{kN-2} \a_r tr(\qt \phi_k^r q) + \!\!\sum_{\tiny{\begin{array}{c}1\leq i \leq k\\2\leq j \leq N+1\\(i,j)\neq(1,N+1) \end{array}}}\!\!\! \b_{i,j} tr(\phi_i^{\,j}) \ee
The Lagrangian dual to $M5$ branes on a sphere with an irregular and a minimal puncture is a $\CN=1$ quiver very similar to (\ref{QUIVER}). This class includes as a special case the $(A_1,D_{2N})$ models found in \cite{Agarwal:2016pjo}. 

The naive superpotential needs to be modified in two ways, using the prescriptions of \cite{Benvenuti:2017lle}: some terms should be dropped from the superpotential due to chiral ring stability, and some gauge singlet fields $\b_{i,j}$ must be added in order to remove operators violating the unitarity bound from the chiral ring.  After these modifications, we have a consistent and complete Lagrangian (\ref{superpot}), so it is possible to proceed with the analysis of the theories. We show that the chiral ring generators of the $\CN=1$ quiver theory precisely map to the generators of the $(A_k,A_{kN+N-1})$ AD models. In particular, non trivial "extended dressed baryons" map to the Higgs Branch of the $\CN=2$ CFT's. As opposed to the case of $(A_1,A_{2N-1})$ discussed in \cite{Benvenuti:2017lle,Benvenuti:2017kud}, in these more general models, some of the $\b$-fields are generators of the chiral ring: without adding the $\b$-fields (i.e. simply stating that some set of $tr(\phi^j)$-operators decouple) it would not be possible to even see the right number of Higgs Branch generators. Moreover, we show that there are non-trivial holomorphic operators in the quiver that cannot take a vacuum expectation value, and map to the $\CN=2$ superpartners of the Coulomb Branch generators, as in \cite{Benvenuti:2017lle,Benvenuti:2017kud} (see also \cite{Giacomelli:2014rna}). 

The consistency of the whole picture is a non trivial check of the prescription of \cite{Benvenuti:2017lle} to add a flipping singlet $\b_{\CO}$ for each unitarity-bound-violating operator $\CO$.

\vspace{2cm}

The paper is organized as follows.

\vspace{0.1cm}

In section \ref{SC} we analyze the $SU(2)$ $4d$ theory dual to $(A_1,D_{4})$ in detail, since it displays some new features with respect to \cite{Benvenuti:2017lle,Benvenuti:2017kud} that will be present also for the Lagrangian theories we find in this paper. We then study the $3d$ mirror RG flow for quiver theories which, through sequential confinement, land on a $\CN=4$ theory with more than one Abelian gauge group. 

In section \ref{AkAn} we discuss the mirror pair $\CT_{UV} \leftrightarrow \tilde{\CT}_{UV}$ that uplifts to the $4d$ Lagrangian (\ref{QUIVER}), (\ref{superpot}) for the theories obtained wrapping $M5$'s on a sphere with an irregular puncture. We study the conformal manifold, the Coulomb branch and the Higgs branch generators, and the superpartners of the Coulomb branch generators, finding a perfect match with the $\CN=2$ CFT's. 

In section \ref{AkAnplusmin} we discuss, with a bit less details, the $4d$ Lagrangian for the theories obtained wrapping $M5$'s on a sphere with one irregular and one minimal puncture. 

In section \ref{3D} we provide a detailed study of the chiral ring and moduli space of the lagrangian description of $(A_1,D_4)$ theory in $3d$: we discuss the dual abelian description of the theory and study in depth the deformations of the theory, recovering as a byproduct the duality between $SU(2)$ adjoint SQCD with one flavor and SQED with two flavors studied in \cite{Benvenuti:2017lle}.

\vspace{0.5cm}
\paragraph{Notation: Quiver diagrams}
\begin{itemize}
\item a circle node $\bpic \node[circle,draw] at(0,-0.3) {$\tiny{\!\!N\!\!}$}; \epic$ denotes a $U(N)$ gauge group;
\item a double-circle node $\bpic \node[circle,draw] at(0,-0.3) {$\tiny{\!\!N\!\!}$}; \node[circle,draw] at(0,-0.3) {$\quad\,$}; \epic$ denotes a $SU(N)$ gauge group;
\item a square node $\bpic \node[rectangle,draw] at(0,-0.3) {$\,\tiny{N}\,$}; \epic$ denotes a $U(N)$ or $SU(N)$ flavor group;
\item sometimes we use an $8$-supercharges notation {\tiny $\bpic \node[circle,draw](x) at(0,-0.3) {$\tiny{\!\!N_1\!\!}$}; \node[circle,draw](y) at(0.9,-0.3) {$\tiny{\!\!N_2\!\!}$}; \draw[-] (x) to (y); \epic$}, links are bifundamental hypers and adjoints in the vector multiplets are implicit;
\item sometimes we use a $4$-supercharges notation {\tiny$\bpic \node[circle,draw](x) at(0,-0.3) {$\tiny{\!\!N_1\!\!}$}; \node[circle,draw](y) at(1.1,-0.3) {$\tiny{\!\!N_2\!\!}$}; \draw [->] (x) to[bend left] (y); \draw [<-] (x) to[bend right] (y); \draw [->] (x) to[out=-45, in=0] (0,-0.8) to[out=180,in=225] (x);\epic$}, arrows are bifundamental or adjoint chiral fields.\end{itemize}

\paragraph{Notation: Flips}
A gauge singlet chiral field $\sigma$ \emph{flips an operator} $\CO$ when it enters the superpotential through the term $\sigma\cdot \CO$. As in \cite{Benvenuti:2017kud}, in this paper we consistently use different names for three classes of flipping fields:
 $\a_r$ fields flip dressed mesons operators, $\b_j$ fields flip $\Tr(\phi^j)$, $\gamma_N$ fields are generated  in the mirror quiver when gauge nodes confine. 

\vspace{0.3cm}
\emph{Note added}: after this work was completed and reported at various talks, we learned about the upcoming paper \cite{Agarwal:2017roi}, which overlaps with sections \ref{AkAn} and \ref{AkAnplusmin} of the present work.

\section{$3d$ mirrors: sequential confinement to $\CN=4$ Abelian quivers}\label{SC}
In this section we study in detail the $3d$ mirror RG flows, which represent our guiding principle for the rest of the paper. We discuss several models which, through sequential confinement, land on $\CN=4$ theories with more than one Abelian gauge group. This generalizes \cite{Benvenuti:2017kud}, in which the mirror RG flow landed on $U(1)$ with $N$ flavors. We discuss in detail the superpotential. We first focus on the $(A_1,D_4)$ theory, an example that illustrates all the basic features of the general case. We then generalize to $(A_1,D_{2N})$  and $(A_k,A_k)$.

We start the section discussing the $4d$ theory dual to $(A_1,D_{4})$ , since it presents some new features with respect to \cite{Benvenuti:2017lle,Benvenuti:2017kud} that will be present also for the two classes of $\CN=1$ Lagrangians found in this paper. The reader interested only in the $4d$ Lagrangians can skip the remaining of this section and look at sections \ref{AkAn} and \ref{AkAnplusmin}.

\subsection{$(A_1,D_{4})$ in $4$ dimensions}
As discovered in \cite{Maruyoshi:2016tqk, Maruyoshi:2016aim,Agarwal:2016pjo}, $4d$ $\CN=2$ $SU(2)$ SQCD with four flavors, upon coupling the moment map to a $\CN=1$ chiral field $A$ and giving next-to-maximal nilpotent vev to $A$, flows in the IR to an $\CN=1$ $SU(2)$ theory which turns out to be dual to the AD model $(A_1,D_{4})$.

We first need to reformulate the Lagrangian as in \cite{Benvenuti:2017lle,Benvenuti:2017kud}, dropping some superpotential terms in order to satisfy chiral ring stability (see also \cite{Collins:2016icw}) and adding a gauge singlet field $\b_2$ in order to decouple the operator $tr(\phi^2)$ that would violate the unitarity bound \cite{Kutasov:2003iy}. 

The consistent and complete theory ha superpotential
\be\label{flipped1}\W= tr(\tilde{b}\phi b) + \alpha_0 tr(\tilde{q}q)+\beta_2\Tr\phi^2.\ee 

As opposed to the cases of cases of adjoint-$SU(N)$ with one flavor studied in \cite{Benvenuti:2017lle,Benvenuti:2017kud}, in this case with two flavors the field $\b_2$ can take a non zero vacuum expectation value (giving a vev to $\b_2$ makes the adjoint $\phi$ massive, reducing the theory to $SU(2)$ SQCD with two flavors, which indeed has a vacuum). So $\b_2$ is a non trivial chiral ring generator and its presence will turn out to be crucial in order to have a complete map to the chiral ring generators of $(A_1,D_4)$.  There are three non-anomalous $U(1)$ global symmetries plus the R-symmetry. This fact also implies that when compactifying to $3d$, in the case with two flavors, a monopole superpotential is generated. The global symmetry charges of the elementary fields and of the chiral ring generators are
\be\label{char212}
\begin{array}{c|cccc}
 &  U(1)_{R} & U(1)_T & U(1)_{B_1} & U(1)_{B_2} \\ \hline
\phi & \frac{1}{3} & \frac{1}{3} &0 &0 \\
q,\qt & \frac{1}{2} & - \frac{1}{2} &\pm 1 & 0\\
b,\bt & \frac{5}{6} & -\frac{1}{6}&  0 & \pm 1 \\ \hline
\a_0 & 1 & 1 & 0 & 0 \\ 
tr(\bt b) & \frac{5}{3} & \frac{1}{3} & 0 & 0 \\ \hline
\CB,\tilde{\CB}=\varepsilon \, q (\phi q) , \varepsilon \, (\qt \phi) \qt & \frac{4}{3} & -\frac{2}{3} & \pm 2 & 0 \\
\CC,\tilde{\CC}=\varepsilon \, b q , \varepsilon \, \bt \qt & \frac{4}{3} & -\frac{2}{3} & 1 & \pm 1 \\ 
\CN,\tilde{\CN}= tr(\bt q), tr(\qt b)  & \frac{4}{3} & -\frac{2}{3} & -1 & \pm 1 \\ 
\CM=tr(\qt \phi q)  &\frac{4}{3} & -\frac{2}{3} & 0 & 0 \\
\b_2  & \frac{4}{3} & -\frac{2}{3} & 0 & 0 \\
\end{array}
\ee
We chose a normalization of $U(1)_T$ such that all the gauge invariant operators that are mapped to the Coulomb (Higgs) branch of the AD model satisfy $R=T$ ($R=-2T$). $\a_0$ is mapped to the CB generator of $(A_1,D_{4})$. In \cite{Benvenuti:2017kud} (see section (2.1.1)), we pointed out that all the gauge invariant operators $\CO_{CB}$ which map to CB operators of the AD model have a superpartner under the hidden supersymmetries $\CO'_{CB}$ with scaling dimension $\Delta[\CO'_{CB}]=\Delta[\CO_{CB}]+1$. The expectation value of $\CO'_{CB}$ is zero at every point of the moduli space. CB generators $\CO_{CB}$ and their superpartners $\CO'_{CB}$ together form the half-BPS $\CN=2$ "Coulomb Branch supermultiplets", that in the Dolan-Osborn notation \cite{Dolan:2002zh} are called $\mathcal{E}_{(R_{\CN=2},0,0)}$. For the $SU(2)$ theory we are discussing\footnote{More generally, the $SU(N)$ adjoint SQCD with $2$ flavors $q,\qt$, $b,\bt$ and \be \CW= tr(\bt \phi b) + \sum_{r=0}^{N-2} \a_r tr(\qt \phi^r q) + \sum_{j=2}^N \b_j tr(\phi^j) \ee is dual to $(A_1,D_{2N})$ AD. The superpartners of the $N-1$ $\a_r$'s are the $N-2$ $\b_j$'s ($j=2,3,\ldots,N-1$) plus $tr(\bt b)$. $\b_{N}$ and the dressed mesons/baryons map to the Higgs Branch of $(A_1,D_{2N})$ AD.} the superpartner of $\a_0$ is the operator $tr(\bt b)$:
\be \a_0 \underleftrightarrow{\quad \CN=2 \quad} tr(\bt b)\ee
 $tr(\bt b)$ cannot take a vev (deriving $\CW$ w.r.t. to $\phi^a_{\,c}$ and contracting with $\bt^a b_c$ we get $tr(\bt b)^2=-\b_2 tr(\bt \phi b)=0$, where the last equality follows from the $\CF$-terms of $b$ or $\bt$). Indeed $R[tr(\bt b)]=R[\a_0]+\frac{2}{3}$, so $\Delta[\CO'_{CB}]=\Delta[\CO_{CB}]+1$.

The other $8$ operators listed (which can be called "dressed baryons" and "dressed mesons") satisfy $R=-2T=\frac{4}{3}$ and we claim are mapped to the HB generators of $(A_1,D_{4})$, which transform in the adjoint of a global $SU(3)$ symmetry, which is the enhancement of our 2 baryonic symmetries $U(1)_B$. Notice that the singlet $\b_2$ we added (using the prescription of \cite{Benvenuti:2017lle}) is part of the octet.

What are the relations satisfied from the eight operators $(\CB,\tilde{\CB},\CC,\tilde{\CC},\CN,\tilde{\CN},\CM,\b_2)$? The result is that the relations are \emph{different} from the relations of the Higgs Branch of $(A_1,D_{4})$. We discuss this issue in detail in section \ref{3D}. Compactifying our theory to $3d$, a monopole superpotential is generated, similarly to \cite{Nii:2014jsa} (the monopole has four fermion zero modes and thanks to the term $\beta_2\Tr\phi^2$ we can soak two of them, obtaining the superpotential term $\beta_2\M$), and the $3d$ compactified theory is dual to $U(1)$ with $3$ flavors $\CN=4$. Upon dropping by hand the monopole superpotential term,  the resulting $3d$ theory is dual to $U(1)$ with $3$ flavors $\CN=2$ with a peculiar superpotential. For the latter duality, in section \ref{3D} we will map the chiral ring generators and also the chiral ring relations. From here, a relevant deformation then takes us to the Abelianization duality with $U(1)$ with $3$ flavors $\CN=4$.

Coming back to $4d$, we interpret the discrepancy between the chiral ring relations in the $\CN=1$ adjoint-$SU(2)$ theory and $(A_1,D_{4})$ as follows: the actual chiral ring is subject to quantum corrections, and the quantum modified chiral ring is precisely the $(A_1,D_{4})$ chiral ring displaying $SU(3)$ global symmetry. It would be important to study this issue in more detail.

\subsection{The $3d$ mirror RG flow: $(A_1,D_4)$}\label{revmirror}
We now study the $3d$ mirror RG flow that lands on the $3d$ mirror of $(A_1,D_4)$, which displays sequential confinement as in \cite{Benvenuti:2017kud}.

In the IR we want to get the dimensional reduction of \ref{flipped1}, which contains a monopole superpotential
\be\label{flippedir}\W_{IR,3d}= tr(\tilde{b}\phi b) + \alpha_0 tr(\tilde{q}q)+\beta_2 (\Tr\phi^2 + \M).\ee 

Notice that $\b_2$ is part of the chiral ring of (\ref{flippedir}) and can have a vev, as was mentioned before, both in four and three dimensions. 

We thus start in the UV from the $3d$ theory
\be\label{flipped}\W_{UV,3d}=\sum_{i=1}^4\tilde{q}_i\phi q_i+\tilde{q}_1q^{2}+\tilde{q}_2q^3+\alpha_0\tilde{q}_3q^1+\beta_2(\Tr\phi^2 + \M),\ee 
which upon integrating out the massive flavors becomes \ref{flippedir}. We call  (\ref{flipped}) and (\ref{flippedir}) $\mathcal{T}'_{3d,UV}$ and  $\mathcal{T}'_{3d,IR}$ respectively. 


We now move to the mirror side. The mirror of $\CN=4$ $SU(2)$ SQCD with four flavors is \cite{Intriligator:1996ex}
\be 
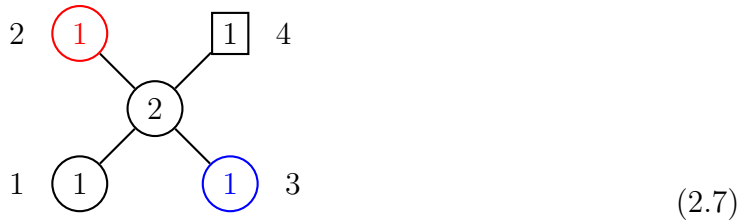
\label{su2mirror}
\begin{tikzpicture}[->, thick]
\node[shape=circle, draw] (1) at (0,0) {$1$};
\node[shape=circle, draw] (2) at (1,1) {$2$};
\node[shape=circle, draw, blue] (3) at (2,0) {$1$};
\node[shape=circle, draw, red] (4) at (0,2) {$1$};  
\node[shape=rectangle, draw] (5) at (2,2) {$1$};
\node[] (6) [left= .2cm of 1] {$1$};
\node[] (7) [left= .2cm of 4] {$2$};
\node[] (8) [right= .2cm of 3] {$3$};
\node[] (9) [right= .2cm of 5] {$4$};
\draw[-] (1) -- (2);
\draw[-] (2) -- (3);
\draw[-] (2) -- (4);
\draw[-] (5) -- (2);
\end{tikzpicture} \ee

For the ease of exposition we numbered the abelian groups in the picture and from now on we will call $b_i,\tilde{b}_i$ the $U(2)\times U(1)_i$ bifundamentals. Since all matter fields transform in the bifundamental representation one abelian gauge factor is redundant and can be dropped. We choose to decouple the vector multiplet $U(1)_4$, so the cartan subgroup of the $SO(8)$ global symmetry of the theory is described by the topological symmetries of the four remaining nodes. In the following we will only need to consider the $SU(3)$ symmetry associated with the nodes $U(1)_1$ and $U(2)$. The singlets in the abelian vector multiplets will be denoted $\varphi_i$ ($i=1,2,3$) whereas the trace and traceless parts of the $U(2)$ adjoint are $\hat{\phi}_2$ and $\phi_2$ respectively. The operators $\Tr\phi^2$ and the monopole of $SU(2)$ SQCD are mapped on the mirror side to $\tilde{b}_4b_3\tilde{b}_3b_4$ and $\tilde{b}_4b_3\tilde{b}_3b_4+\tilde{b}_2b_3\tilde{b}_3b_2$ respectively.

The $SO(8)$ global symmetry of SQCD arises quantum mechanically in the mirror theory: the Cartan subgroup $U(1)^4$ corresponds to the topological symmetry associated with the four abelian gauge groups in Figure \ref{su2mirror}, whereas the other generators are related to monopole operators of dimension one, whose multiplets contain conserved currents (\cite{Gaiotto:2008ak}). In the rest of the paper we only need to consider the $SU(3)$ subgroup associated with the gauge groups $U(1)_1$ and $U(2)$. The map between off-diagonal components of the meson and monopoles is as follows: 
\be\label{meson}\left(\begin{array}{ccc}
 & \tilde{q}_1q^2 & \tilde{q}_1q^3 \\
 \tilde{q}_2q^1 & & \tilde{q}_2q^3 \\
 \tilde{q}_3q^1 & \tilde{q}_3q^2 & \\
\end{array}\right)
\leftrightarrow \left(\begin{array}{ccc}
 & \M^{+0} & \M^{++} \\
 \M^{-0} & & \M^{0+} \\
 \M^{--} & \M^{0-} & \\
\end{array}\right)\ee
The two Cartan components of the meson matrix are mapped to $\varphi_1$ and $\hat{\phi}_2$.
In (\ref{meson}) we have included only the charges under the topological symmetries related to $U(1)_1$ and $U(2)$, the others being trivial. 

\subsubsection*{The mirror of $\CT'_{3d,UV}$ and $(A_1,D_4)$ AD theory}

Using the results reviewed before, we find that the mirror dual of $\mathcal{T}'_{3d,UV}$ is the gauge theory in (\ref{su2mirror}) with superpotential  
\beqa\label{mirror2}\W \nn &=& \sum_i\varphi_i\tilde{b}_ib^i-\hat{\phi}_2(\sum_i\tilde{b}_ib^i)- \Tr(\phi_2(\sum_ib^i\tilde{b}_i))+\M^{+,0}+\M^{0,+}+\alpha_0\M^{--}\\ &&+\beta_2(\tilde{b}_4b_3\tilde{b}_3b_4+2\tilde{b}_2b_3\tilde{b}_3b_2).\eeqa 
where we included the dynamically generated monopole term. 

We now use the monopole duality discovered in \cite{Benini:2017dud} (see also \cite{Benvenuti:2016wet, Collinucci:2016hpz} for previous Abelian examples and \cite{Amariti:2017gsm} for a brane interpretation), which in the case of interest to us states that a $3d$ $\CN=2$ theory $U(N_f-1)$ with $N_f$ flavors and $\CW=\M^+$ is dual to a Wess-Zumino model with superpotential
\be \CW= \gamma_{N_f} det(X_{N_f}) \ee

Before proceeding, we would like to remark that in what follows the two numerical coefficients of the $\b_2$ term in (\ref{mirror2})  could be replaced by two arbitrary numbers without affecting the final result. The important point is that the $\b_2$ term in (\ref{mirror2}) is not proportional to $\beta_2\tilde{b}_4b_3\tilde{b}_3b_4$, as we would get if in (\ref{mirror2}) we neglected the mirror of the superpotential term $\b_2\M$.

\begin{figure}[ht!]
\centering
\begin{tikzpicture}[->, thick]
\node[shape=circle, draw] (2) at (0,0) {$2$};
\node[shape=circle, draw, blue] (3) at (1,-1) {$1$};
\node[shape=circle, draw, red] (4) at (-1,1) {$1$};  
\node[shape=rectangle, draw] (5) at (1,1) {$1$};
\node[] (7) [left= .2cm of 4] {$2$};
\node[] (8) [right= .2cm of 3] {$3$};
\node[] (9) [right= .2cm of 5] {$4$};

\draw[-] (2) -- (3);
\draw[-] (2) -- (4);
\draw[-] (5) -- (2);
;

\end{tikzpicture}
\caption{Mirror of $\mathcal{T}'_{3d,UV}$ after confinement of the $U(1)_1$ gauge group.}\label{sqcd22}
\end{figure}
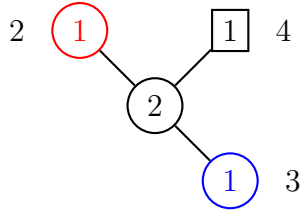

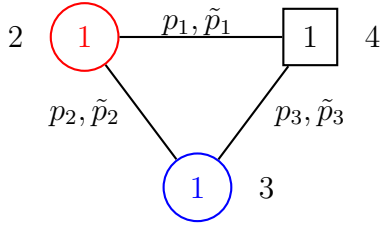
\begin{figure}[ht!]
\centering
 \begin{tikzpicture}[->, thick]
\node[shape=circle, draw, blue, minimum height=.9cm] (1) at (0,-1) {$1$};
\node[shape=rectangle, draw, minimum height=.75cm] (2) at (1.5,1) {$\;1\;$};
\node[shape=circle, draw, red, minimum height=.9cm] (3) at (-1.5,1) {$1$};
\node[] (4) [left= .2cm of 3] {$2$};
\node[] (5) [right= .2cm of 1] {$3$};
\node[] (6) [right= .2cm of 2] {$4$};
\node[] (7) at (0,1.2) {$p_1,\tilde{p}_1$};
\node[] (8) at (-1.5,0) {$p_2,\tilde{p}_2$};
\node[] (9) at (1.5,0) {$p_3,\tilde{p}_3$};

\draw[-] (1) -- (2);
\draw[-] (1) -- (3);
\draw[-] (2) -- (3);
; 
\end{tikzpicture}
\caption{Mirror of $\mathcal{T}'_{3d,UV}$ after confinement of the $U(2)$ gauge group.}\label{sqcd23}
\end{figure}

According to the monopole duality, the gauge group $U(1)_1$ confines leaving behind the $U(2)$ adjoint chiral $X_2$. 
The theory (\ref{mirror2}) now becomes as in Figure \ref{sqcd22}

\noindent with superpotential 
\bea\label{mirror22}\nn \W &=&\varphi_1\Tr X_2+\sum_{i>1}\varphi_i\tilde{b}_ib^i-\hat{\phi}_2\left(\Tr X_2+\sum_{i>1}\tilde{b}_ib^i\right) - \Tr\left[\phi_2\left(X_2 +\sum_{i>1}b^i\tilde{b}_i\right)\right]\\ 
&& +\gamma_2\det X_2+\M^{+}+\alpha_0 \M^{-}+\beta_2(\tilde{b}_4b_3\tilde{b}_3b_4+2\tilde{b}_2b_3\tilde{b}_3b_2).\eea 
In this formula $\M^{\pm}$ denote the monopoles charged under the topological symmetry associated with the $U(2)$ gauge node. As is clearly displayed by the superpotential, $X_2$ and $\phi_2$ become massive and can be integrated out. 
At this stage the $U(2)$ gauge group has three flavors and no adjoint matter, so according to the monopole duality it confines and is traded for a $3\times3$ chiral multiplet $X_3$, which is nothing but the dual of $\tilde{b}_ib_j$ ($i,j=2,3,4$). This also generates the superpotential term $\gamma_3\det X_3$. Notice that the equations of motion of (\ref{mirror22}) impose the constraint $X_2=-\sum_{i>1}b^i\tilde{b}_i$. Using this fact we can express $\det X_2$ in terms of traces of $X_3$: 
\be\label{id2}\det X_2=\frac{(\Tr X_2)^2-\Tr X_2^2}{2}=\frac{(\tilde{b}_ib^i)^2-\Tr((\tilde{b}_ib_j)^2)}{2}=\frac{(\Tr X_3)^2-\Tr X_3^2}{2}.\ee
Notice that in theory (\ref{sqcd22}) the cartan subgroup of the $U(3)$ symmetry under which $\tilde{b}_ib_j$ ($i,j=2,3,4$) transforms in the adjoint representation is gauged: the $U(1)_{2,3,4}$ symmetries are generated respectively by the $3\times3$ matrices $\text{diag}(1,0,0)$, $\text{diag}(0,0,1)$ and $\text{diag}(0,1,0)$. Our convention will be that these groups act in the same way on the matrix $X_3$ after confinement of the $U(2)$ gauge group. As a result, the off-diagonal components of $X_3$ become bifundamental hypermultiplets charged under the leftover $U(1)_i$ symmetries and we relabel the fields as follows:
$$(X_3)_1^2,(X_3)_2^1\leftrightarrow p_1,\tilde{p}_1;\quad (X_3)_1^3,(X_3)_3^1\leftrightarrow p_2,\tilde{p}_2;\quad (X_3)_2^3,(X_3)_3^2\leftrightarrow p_3,\tilde{p}_3.$$
After confinement of the $U(2)$ gauge group the theory in Figure \ref{sqcd22} becomes as in Figure \ref{sqcd23} 

\noindent The fields $\varphi_i$ now only appear in the superpotential terms 
\be\W=(\varphi_2-\hat{\phi}_2)(X_3)_1^1+\hat{\phi}_2(X_3)_2^2+(\varphi_3-\hat{\phi}_2)(X_3)_3^3\dots\ee 
As a consequence they become massive and their $\CF$-terms set to zero the diagonal components of $X_3$. The remaining fields are $\alpha_0$, $\beta_2$, $\gamma_{2,3}$ and $p_i,\tilde{p}_i$ with superpotential 
\be\label{mirror23}\W=-\frac{\gamma_2}{2}\sum_i\tilde{p}_ip_i+\beta_2(\tilde{p}_2p_2+2\tilde{p}_3p_3)+\gamma_3\det X_3+\alpha_0\gamma_3.\ee 
$\alpha_0$ and $\gamma_3$ are massive, so the last two terms disappear, leaving us with 
\be\label{mirror24}\W=-\frac{\gamma_2}{2}\sum_i\tilde{p}_ip_i+\beta_2(\tilde{p}_2p_2+2\tilde{p}_3p_3),\ee 
which is equivalent to an $\CN=4$ superpotential. This model is known to be the mirror of $\CN=4$ SQED with three flavors and is precisely the mirror of $(A_1,D_4)$ Argyres-Douglas theory proposed by Nanopoulos and Xie \cite{Nanopoulos:2010bv}. This model
is the mirror of $\CN=4$ SQED with three flavors, which must then be the Abelianization of the $3d$ reduction of (\ref{flipped1}). We will check this statement explicitly in section \ref{3D}.\footnote{Notice that if we had neglected the mirror of the monopole term in (\ref{mirror2}), instead of (\ref{mirror24}) we would have found 
\be\label{mirror241}\W=-\frac{\gamma_2}{2}\sum_i\tilde{p}_ip_i+\beta_2\tilde{p}_3p_3.\ee 
This is the supepotantial of the mirror dual of $3d$ $\CN=2$ SQED with three flavors $Q_1, Q_2, p$, one singlet $\Phi$ and superpotential 
\be\W=\Phi(\Qt_1Q_1+\Qt_2Q_2).\ee
We will test this statement in section \ref{strangeU1}.}

\subsection{Flow to the mirror of $(A_1,D_6)$} 

The $(A_1,D_6)$ AD theory can be obtained by starting from $SU(3)$ SQCD with 6 flavors, whose mirror is the quiver 
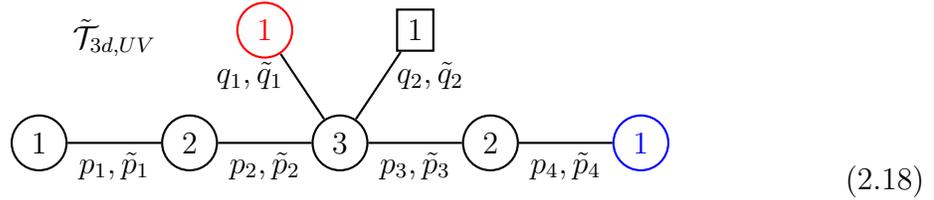
\be \label{sqcd3} \bpic[->, thick]
  \path (2,1.4) node {$\tilde{\CT}_{3d,UV}$} (1,0) node[circle,draw](x1) {$1$} -- (3,0) node[circle,draw](x2) {$2$} -- (5,0) node[circle,draw](x5) {$3$} -- (4,1.5) node[circle,draw,red,thick](x6) {$1$} -- (6,1.5) node[rectangle,draw,thick](x7) {$1$}  --  (7,0) node[circle,draw](x9) {$2$} -- (9,0) node[circle,draw,blue](x10) {$1$}; 
\draw [-] (x1) to (x2);
\draw [-] (x2) to (x5);
\draw [-] (x5) to (x6);
\draw [-] (x5) to (x7);
\draw [-] (x5) to (x9);
\draw [-] (x9) to (x10);
\node at (3.8,0.9) {$q_1,\qt_1$};
\node at (6.2,0.9) {$q_2,\qt_2$};
\node at (6,-0.3) {$p_3,\pt_3$};
\node at (8,-0.3) {$p_4,\pt_4$};
\node at (4,-0.3) {$p_2,\pt_2$};
\node at (2,-0.3) {$p_1,\pt_1$};
  \epic
\ee 
 and adding a $5\times5$ flipping field, to which we give a maximal nilpotent vev. In the IR we are left with 4 flipping singlets but two of them violate the unitarity bound and decouple; the same happens to $\Tr\Phi^2$ and $\Tr\Phi^3$ and the two ``surviving'' flipping fields are interpreted as the Coulomb branch operators of $D_6$ AD theory. 

As we did in the previous case, we will use a slightly different definition of this theory in which we don't have to decouple any operator: we start from $SU(3)$ SQCD with six flavors, turn on four off-diagonal mass terms and flip the operators $\Tr\Phi^2$ and $\Tr\Phi^3$. We also introduce the two flipping fields which do not decouple in the IR. The superpotential of our UV theory is 
\be\label{uvd6}\W=\sum_{i=1}^6\tilde{q}_i\Phi q^i+\beta_2\Tr\Phi^2+\beta_3\Tr\Phi^3+\sum_{i=1}^4\tilde{q}_iq^{i+1} +\alpha_{0}\tilde{q}_5q^1+\alpha_1(\tilde{q}_4q^1+\tilde{q}_5q^2)+\beta_3\{\M\phi\}.\ee
In the above formula we have included the superpotential term involving the dressed monopole generated in the compactification to 3d \cite{Nii:2014jsa}. This is similar to the $SU(2)$ case discussed previously. Once we have integrated out the massive flavors, we are left in the IR with $SU(3)$ SQCD with two flavors ($q$ and $b$) and superpotential 
\be\label{irgauge}\W=\tilde{b}\Phi b + \alpha_0\tilde{q}q + 2\alpha_1\tilde{q}\Phi q + \beta_2\Tr\Phi^2 + \beta_3\Tr\Phi^3+\beta_3\{\M\phi\}.\ee 
This theory is IR equivalent to $D_6$ AD and there are no unitarity bound violations. 

We claim that the dimensional reduction of the UV theory described in the previous paragraph flows in the IR to the dimensional reduction of $D_6$ AD theory. Let us now discuss the mirror of the UV theory: the four mass terms are mapped to superpotential terms involving monopole operators charged under a single topological $U(1)$ group, one for each gauge node in the two tails of (\ref{sqcd3}) except the abelian node on the right (we call them $\M_i^+$ for $i=1,2,3,4$), whereas the terms involving the singlets $\alpha_0$ and $\alpha_1$ appearing in {\ref{irgauge}} are mapped to $\alpha_1(\M^{---0}+\M^{0---})$ and $\alpha_0\M^{----}$, the monopoles with charge $-1$ under the topological symmetries associated with $U(1)_L,U(2)_L,U(3)$, $U(2)_L,U(3),U(2)_R$ and $U(1)_L,U(2)_L,U(3),U(2)_R$ respectively. We conclude that the mirror of our UV theory is the quiver (\ref{sqcd3}) with superpotential
\be\label{mirror4}\begin{array}{ll}\W=&\W_{\CN=4}+\sum_i\M_i^+ + \alpha_1(\M^{---0}+\M^{0---}) + \alpha_0\M^{----}\\ &+ \beta_2q_2p_3\tilde{p}_3\tilde{q}_2+\beta_3(2q_2p_3p_4\tilde{p}_4\tilde{p}_3\tilde{q}_2 +q_1p_3p_4\tilde{p}_4\tilde{p}_3\tilde{q}_1).\end{array}\ee 
The analysis proceeds as in section 3 of \cite{Benvenuti:2017kud} until we dualize the $U(3)$ node, leaving us with the theory 
\be\label{stepnew} 
 \begin{tikzpicture}[->, thick]
\node[shape=circle, draw, minimum height=.8cm,blue] (1) at (4,0) {$1$};
\node[shape=circle, draw, minimum height=.8cm] (2) at (2,0) {$2$};
\node[shape=circle, draw, minimum height=.8cm,red,thick] (3) at (0.9,1.5) {$1$};
\node[shape=rectangle, draw, minimum height=.8cm,thick] (4) at (3.1,1.5) {$\,\,1\,\,$};
\node[] (5) at (3,-0.3) {$p_4,\tilde{p}_4$};
\node[] (6) at (0.9,0.7) {$v,\tilde{v}$}; 
\node[] (7) at (3.1,0.7) {$w,\tilde{w}$}; 
\node[] (8) at (2,1.8) {$Q_1\!,\!\Qt_1$};
\draw[-] (1) -- (2);
\draw[-] (2) -- (3);
\draw[-] (2) -- (4);
\draw[-] (4) -- (3);
\end{tikzpicture}
\ee
and superpotential 
\beqa\W \nn &=&\gamma_2(...)+\gamma_3(...)+\gamma_4(...)+\M^{+} + \alpha_1(\gamma_4+M') + \alpha_0\M^{-} + \beta_2\tilde{w}w\\ && + \beta_3(2\tilde{w}p_4\tilde{p}_4w+\tilde{v}p_4\tilde{p}_4v)+\varphi_5\tilde{p}_4p_4.\eeqa 
The monopoles appearing in this formula are charged under the topological symmetry of the $U(2)$ node, $M'$ is the dual of $\M^{0---}$ appearing in (\ref{mirror4}) and $\varphi_5$ denotes the chiral in the $U(1)_R$ vectormultiplet. The first three terms are the superpotential terms generated dynamically when we dualize the nodes and have the following form \cite{Benvenuti:2017kud}: 
\be\label{det2}-\gamma_2(\Qt_1Q_1+\tilde{v}v+\tilde{w}w)\ee 
\be\label{det3}\gamma_3(Q_1\tilde{w}v+\tilde{Q}_1\tilde{v}w+\tilde{v}p_4\tilde{p}_4v+\tilde{w}p_4\tilde{p}_4w)\ee 
\be\label{det4}\gamma_4(\tilde{v}v\tilde{w}w-\tilde{w}v\tilde{v}w-\tilde{Q}_1\tilde{v}p_4\tilde{p}_4w-Q_1\tilde{w}p_4\tilde{p}_4v)\ee

Now the $U(2)$ node confines leaving behind a $3\times3$ meson $N_{ij}$ 
\be\label{expr41} N=\left(\begin{array}{ccc}
N_{11} & Q_2 & N_{13} \\
\Qt_2 & N_{22} & N_{23} \\
N_{31} & N_{32} & N_{33}
        \end{array}\right)
\ee 
which provides one extra bifundamental ($Q_2$ and $\Qt_2$) of the $U(1)$'s denoted by a square and red circle in (\ref{stepnew}). The field $M'$ is now identified with $\gamma_4$. The fields $\alpha_0$, $\alpha_1$, $\gamma_4$ and $\varphi_5$ become massive and we end up with the theory 
\be \begin{tikzpicture}[->, thick]
\node[shape=circle, draw, minimum height=.8cm,blue] (2) at (2,0) {$1$};
\node[shape=circle, draw, minimum height=.8cm,red,thick] (3) at (0.9,1.5) {$1$};
\node[shape=rectangle, draw, minimum height=.8cm,thick] (4) at (3.1,1.5) {$\,\,1\,\,$};\node[] (8) at (2,1.8) {$Q_{1,2}\!,\!\Qt_{1,2}$};
\draw[-] (2) -- (3);
\draw[-] (2) -- (4);
\draw[-] (4) -- (3);
\end{tikzpicture}
\ee
\be\begin{array}{ll} \W= & -\gamma_2(\tilde{Q}_1Q_1+N_{11}+N_{22})+\gamma_3(\tilde{Q}_1Q_2+\Qt_2Q_1+N_{13}N_{31}+N_{23}N_{32})\\
& +\beta_2N_{22}+\beta_3(2N_{23}N_{32}+N_{13}N_{31}).\end{array}\ee 
The fields $N_{11}$ and $N_{22}$ are now massive and integrating them out we are left with 
\be\W=\gamma_3(\tilde{Q}_1Q_2+\Qt_2Q_1+N_{13}N_{31}+N_{23}N_{32})+\beta_3(2N_{23}N_{32}+N_{13}N_{31}),\ee 
which is equivalent to the superpotential of the $\CN=4$ theory proposed in \cite{Nanopoulos:2010bv}.

\subsection{Generic case $(A_1,D_{2N})$} 
In general, we can obtain the mirror of the $(A_1,D_{2N+2})$ theory starting from the $U(1)$ mirror of $(A_1,A_{2N-1})$ and introducing one extra $U(1)$ node and two hypermultiplets ($v$ and $w$), in the following way 
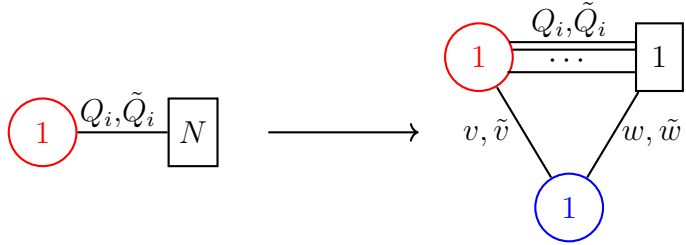
\begin{center}
 \begin{tikzpicture}[->, thick]
\node[shape=circle, draw, minimum height=.9cm,red] (1) at (-1,0) {$1$};
\node[shape=rectangle, draw, minimum height=.9cm] (2) at (1,0) {$N$};

\node[shape=circle, draw, minimum height=.9cm,red] (3) at (4.8,1) {$1$};
\node[shape=rectangle, draw, minimum height=.9cm] (4) at (7.2,1) {$\,1\,$};
\node[shape=circle, draw, minimum height=.9cm,blue] (5) at (6,-1) {$1$};
\node[] (6) at (4.9,0) {$v,\tilde{v}$}; 
\node[] (7) at (7.1,0) {$w,\tilde{w}$};
\node[]  at (-0,0.3) {$Q_{i},\!\Qt_{i}$};
\node[]  at (6,1.5) {$Q_{i},\!\Qt_{i}$};
\node[]  at (6,0.95) {$\cdots$};
\draw[=>] (2,0)--  (4,0);

\draw[-] (1) -- (2); 
\draw[-] (5.2,1.2) -- (6.9,1.2);
\draw[-] (5.25,1.1) -- (6.9,1.1);
\draw[-] (5.18,0.8) -- (6.9,0.8);
\draw[-] (3) -- (5); 
\draw[-] (4) -- (5);
; 
\end{tikzpicture}
\end{center}

As shown in \cite{Benvenuti:2017kud}, the $3d$ mirror of the $(A_1,A_{2N-1})$ theory has superpotential 
\be \CW_{(A_1,A_{2N-1})} = \gamma \, \sum_{i=1}^N Q_i \Qt_{N-i+1} \ee
where $Q_i,\Qt_i$ is the fundamental hypermultiplet generated at the ith-step, dualizing
down the second tail in the sequential confinement. The $U(1) \times U(1)$ $3d$ mirror at the bottom of the RG flow to $(A_1,D_{2N+2})$ has superpotential 
\be \W_{(A_1,D_{2N+2})} =\gamma \left( \sum_{i=1}^N Q_i \Qt_{N-i+1} + \tilde{v}v+\tilde{w}{w} \right)+\beta(\tilde{w}w-\tilde{v}v).\ee 

\subsection{Linear quivers and the $(A_k,A_k)$ theory}
The set of theories $(A_k,A_k)$ lies at the intersection of the two classes of models we are going to consider in the following two sections, setting $N=1$. Here we want to show how the sequential confinement works for this class, i.e. how the correct $\CN=4$ superpotential is recovered in the IR of the mirror RG flow. The $3d$ mirror IR quivers are the complete graphs proposed in \cite{Xie:2012hs}: $k+1$ nodes with a bifundamental for each pair of nodes, where one of the nodes is ungauged: 
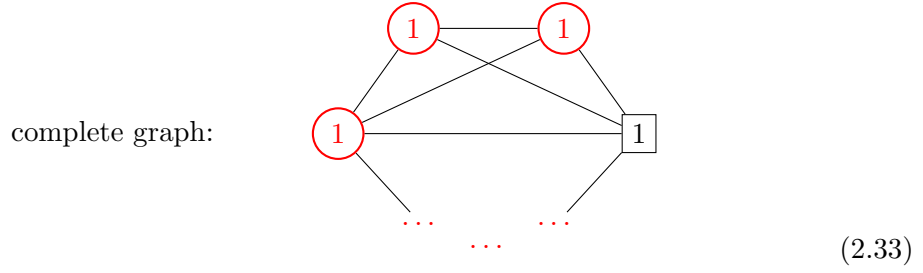
\begin{small}  \be \bpic 
\path (-2,0) node[circle,draw,red,thick](y1) {$1$} -- (-1,1.4) node[circle,draw,red,thick](y2) {$1$} -- (1,1.4) node[circle,draw,red,thick](y3) {$1$} -- (2,0) node[rectangle,draw](y4) {$1$} -- (-0.9,-1.2) node[red] (y5) {$\dots$} -- (0.9,-1.2) node[red] (y6) {$\dots$} -- (0,-1.5) node[red] {$\dots$};
\path (-5,0) node {complete graph:};
\draw [-] (y1) to (y2);\draw [-] (y1) to (y3);\draw [-] (y1) to (y4);\draw [-] (y2) to (y3);\draw [-] (y2) to (y4);\draw [-] (y4) to (y3);\draw [-] (y6) to (y4);\draw [-] (y1) to (y5);
  \epic \ee \end{small}

We focus on the case $k=3$ for simplicity. We start by noticing that in four dimensions the $\CN=2$ $SU(2) \times SU(3)$ gauge theory 
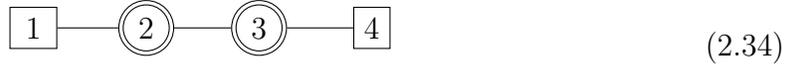
\be \bpic  \path (-3,0) node[rectangle,draw](x0) {$\,1\,$} -- (-1.5,0) node[circle,draw](x1) {$2$} -- (0,0) node[circle,draw](x2) {$3$} -- (1.5,0) node[rectangle,draw](x3) {$4$} ;
\node[circle,draw] at (-1.5,0) {$\!\!\quad$};\node[circle,draw] at (0,0) {$\!\!\quad$};
\draw [-] (x1) to (x0);
\draw [-] (x1) to (x2);
\draw [-] (x2) to (x3);
  \epic \ee
  ows in the IR to the $(A_3,A_3)$ theory provided we add a $4\times4$ flipping field and give it a maximal nilpotent vev which breaks the $SU(4)$ global symmetry completely (see the next section for a detailed discussion about this statement). In the IR the two quadratic casimirs $\Tr\Phi_1^2$, $\Tr\Phi_2^2$ and one of the singlets saturate the unitarity bound and decouple. Following our prescription for the dimensional reduction, we now consider  the theory with superpotential (we call the four $SU(3)$ fundamentals $q_i$)
\be\W=\W_{\CN=4}+\tilde{q}_1q^2+\tilde{q}_2q^3+\tilde{q}_3q^4+\beta_{2,1}\Tr\Phi_1^2+\beta_{2,2}\Tr\Phi_2^2+\alpha_0\tilde{q}_4q^1+\alpha_1(\tilde{q}_3q^1+\tilde{q}_4q^2).\ee 
The mirror of this theory is the quiver 
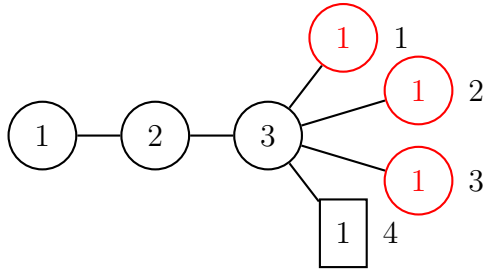
\begin{figure}[ht!]
\begin{center}
 \begin{tikzpicture}[->, thick]
\node[shape=circle, draw, minimum height=.9cm] (1) at (0,0) {$1$};
\node[shape=circle, draw, minimum height=.9cm] (2) at (1.5,0) {$2$};
\node[shape=circle, draw, minimum height=.9cm] (3) at (3,0) {$3$};
\node[shape=circle, draw, minimum height=.9cm,red] (4) at (4,1.3) {$1$};
\node[shape=rectangle, draw, minimum height=.9cm] (5) at (4,-1.3) {$\,1\,$};
\node[shape=circle, draw, minimum height=.9cm,red] (6) at (5,0.6) {$1$};
\node[shape=circle, draw, minimum height=.9cm,red] (7) at (5,-0.6) {$1$};
\node[] (10) [right= .05cm of 4] {$1$};
\node[] (11) [right= .05cm of 6] {$2$};
\node[] (8) [right= .05cm of 7] {$3$};
\node[] (9) [right= .05cm of 5] {$4$};

\draw[-] (1) -- (2);
\draw[-] (2) -- (3);
\draw[-] (3) -- (4);
\draw[-] (3) -- (5);
\draw[-] (3) -- (6);
\draw[-] (3) -- (7);
; 
\end{tikzpicture}
\end{center}
\end{figure}

\noindent with a superpotential of the form 
\beqa\label{mirrorn}\nn \W &=& \W_{\CN=4}+\M^{+00}+\M^{0+0}+\M^{00+}+\alpha_0\M^{---}+\alpha_1(\M^{--0}+\M^{0--})\\ 
&& +\beta_{2,1}\sum_{i,j}\tilde{q}_iq_j\tilde{q}_jq_i+\beta_{2,2}(...)\eeqa
In the above formula we have included the charge of monopole operators under the three topological $U(1)$ symmetries associated with the three gauge nodes $U(1)$, $U(2)$ and $U(3)$. The terms involving $\beta_{2,i}$ include a sum of terms quartic in the bifundamental fields. These are the mirror duals of the operators $\Tr\Phi_1^2$, $\Tr\Phi_2^2$ and the monopole operators which enter in the superpotential terms which arise dynamically. We will not attempt to determine explicitly these terms, although it would be important to fill in this gap. 

We denote the singlets in the $\CN=4$ vector multiplets of the gauge groups $U(1)_i$ ($i=1,2,3$) with $\varphi_i$ and the trace part of the adjoint chiral in the $U(3)$ vectormultiplet as $\varphi_4$.  As usual, one $U(1)$ factor of the gauge group decouples since all matter fields transform in the bifundamental representation and we choose to decouple $U(1)_4$.

Following the usual procedure of sequential confinement, we find that the gauge groups in the tail $U(1)$, $U(2)$ and $U(3)$ 
confine leaving behind a $4\times4$ chiral multiplet N. All the diagonal components of $N$ become massive due to the couplings $\varphi_iN_{ii}$ appearing in the $\CN=4$ part of the superpotential and are set to zero by the $\CF$-terms of the fields $\varphi_i$. The off-diagonal components of $N$ become bifundamentals of the left-over abelian gauge groups, leading to the conclusion that the theory becomes 
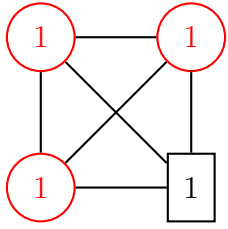
\begin{center}
 \begin{tikzpicture}[->, thick]
\node[shape=circle, draw, minimum height=.9cm,red] (1) at (0,0) {$1$};
\node[shape=circle, draw, minimum height=.9cm,red] (2) at (0,2) {$1$};
\node[shape=circle, draw, minimum height=.9cm,red] (3) at (2,2) {$1$};
\node[shape=rectangle, draw, minimum height=.9cm] (4) at (2,0) {$\,1\,$};

\draw[-] (1) -- (2);
\draw[-] (2) -- (3);
\draw[-] (3) -- (4);
\draw[-] (1) -- (4);
\draw[-] (1) -- (3);
\draw[-] (2) -- (4);
; 
\end{tikzpicture}
\end{center}
Similarly to the $D_6$ theory studied before, we can express all the superpotential terms generated dynamically in terms of the components of the matrix $N$. The same is true for the terms involving $\beta_{2,i}$ fields, which now become bilinear in the components of $N$ (although we don't know their precise form). We find the following superpotential: 
\be\W=-\frac{\gamma_2}{2}\Tr N^2+\frac{\gamma_3}{3}\Tr N^3 + \gamma_4\det N +\alpha_0\gamma_4+2\alpha_1\gamma_3+\beta_{2,1}(\dots)+\beta_{2,2}(\dots).\ee 
In the above formula $\gamma_{3,4}$ are set to zero by the F-terms of $\alpha_0$ and $\alpha_1$ and the superpotential reduces to \be\W=-\gamma_2(N_{12}N_{21}+N_{13}N_{31}+N_{14}N_{41}+N_{24}N_{42}+N_{23}N_{32}+N_{34}N_{43})+\beta_{2,1}(\dots) +\beta_{2,2}(\dots).\ee 
In conclusion, we have come across a theory with precisely the matter content of the $\CN=4$ theory proposed in \cite{Xie:2012hs} as the 3d mirror of $(A_3,A_3)$ AD theory. Furthermore, the superpotential has precisely the form we expect for a theory with eight supercharges: there are singlet chirals in one-to-one correspondence with the vectormultiplets of the theory and they couple to the bifundamental matter fields. This is a strong consistency check for our proposal: the quiver   
\be \bpic  \path (-3,0) node[rectangle,draw](x0) {$\,1\,$} -- (-1.5,0) node[circle,draw](x1) {$2$} -- (0,0) node[circle,draw](x2) {$3$} -- (1.5,0) node[rectangle,draw](x3) {$4$} ;
\node[circle,draw] at (-1.5,0) {$\!\!\quad$};\node[circle,draw] at (0,0) {$\!\!\quad$};
\draw [-] (x1) to (x0);
\draw [-] (x1) to (x2);
\draw [-] (x2) to (x3);
  \epic \ee
(deformed as explained) represents a UV lagrangian completion of the $(A_3,A_3)$ theory. We will perform further checks in the next sections.

For arbitrary $k$, the generalization is as follows: we start from a quiver which is a $T(SU(k+1))$ tail, with all but one flavors at the end of the tail gauged by $U(1)$'s. We introduce $k-1$ singlets $\beta_{2,i}$ flipping the mirrors of all $\Tr\phi_i^2$ operators and $k-1$ singlets $\alpha_i$ flipping the sum of all monopole operators charged under $k-i$ topological $U(1)$ groups. All the nodes in the tail confine and are traded for a $(k+1)\times(k+1)$ chiral multiplet whose diagonal entries are set to zero by $\CF$-terms. The off-diagonal components of this chiral provide bifundamental hypermultiplets charged under all possible pairs of the surviving $U(1)$ gauge groups, leaving us with the complete graph with $k+1$ vertices. 

The superpotential terms involving $\alpha_i$ fields become mass terms and their $\CF$-terms set to zero all the dynamically generated superpotential terms, except the one generated at the first step, when the $U(1)$ node confines. We are then left with the multiplet $\gamma_2$ coupled to the trace of the square of the $(k+1)\times(k+1)$ chiral multiplet. In conclusion, we find the complete graph with $k$ massless singlets coupled to terms quadratic in the bifundamental fields, which is precisely the matter content and superpotential of the $\CN=4$ theory.

\section{$(A_k,A_{kN+N-1})$. $M5$'s on a sphere with an irregular puncture}\label{AkAn}
Armed with our $3d$ mirror "sequential confinement" interpretation of the flow to $(A_1,A_{2N-1})$ AD models \cite{Benvenuti:2017lle,Benvenuti:2017kud}, we are in the position to generalize the story. In this section we find $4d$ Lagrangian field theories that flow to the $(A_k,A_{kN+N-1})$ AD models, for generic $k$ positive. The $(A_k,A_{kN+N-1})$ models can also be given a Gaiotto description in terms of $k$ $M5$ branes wrapping a sphere with an irregular puncture (see the Appendix).

Our strategy is to find UV $3d$ $\CN=4$ mirror pairs $\CT_{UV} \leftrightarrow \tilde{\CT}_{UV}$ such that in both $\CT_{UV} $ and $\tilde{\CT}_{UV}$ the non-Abelian nodes are balanced, i.e. they have $N_f=2N_c$. Upon flipping the Coulomb branch moment map in $\tilde{\CT}_{UV}$, a maximal nilpotent vev sequentially confines $\tilde{\CT}_{UV}$ to a $\CN=4$ Abelian quiver, which is the complete graph $3d$ mirror of the Ad models.

Then we uplift $\CT_{UV} $ to $4d$, so our $4d$ starting point is a $\CN=2$ linear balanced quiver gauge theory where the last node has nontrivial global symmetry (in this case the symmetry will actually be $SU(kN+N)$). Flipping the Higgs Branch moment map and giving a maximal nilpotent vev, we end up with an apparently $\CN=1$ $4d$ IR gauge theory, that is actually dual to $(A_k,A_{kN+N-1})$ AD theory, as expect from the $3d$ mirror arguments.

It is interesting to note that the set of theories $(A_k,A_{kN+N-1})$ (with $k>1$) are special from the S-duality point of view: they are the only ones among the class of $(G,G')$ models (defined in \cite{Cecotti:2010fi}) that displays an infinite dimensional S-duality group \cite{Caorsi:2016ebt}. From our point of view, the set of theories $(A_k,A_{kN+N-1})$ is special since they admit an $\CN=1$ Lagrangian coming from a Maruyoshi-Song deformation of a $\CN=2$ quiver that reduced to $3d$ has a mirror where all non-Abelian gauge groups are balanced. It would be interesting to find a possible relation between these two different perspectives.

\subsection{$3d$ sequential confinement to the complete graph quiver}
In this subsection we qualitatively discuss the $3d$ story. The $\CT_{UV} \leftrightarrow \tilde{\CT}_{UV}$ mirror pair is:

\bea \nn 
\bpic  \path (-3,0) node {$\CT_{3d,UV}:$} (-1.5,0) node[circle,draw](x1) {$N$} -- (0,0) node[circle,draw](x2) {$2N$} -- (1.2,0) node(x3) {$\cdots$} -- (2.7,0) node[circle,draw](x4) {\small$\!\!\!kN\!\!-\!\!N\!\!\!$} -- (4.2,0) node[circle,draw](x5) {$\,k \, N\,$} -- (6.2,0) node[rectangle,draw](x6) {$\!kN\!+\!N\!$} ;
\node[circle,draw] at (-1.5,0) {$\quad$};\node[circle,draw] at (0,0) {$\,\,\quad$};
\node[circle,draw] at (2.7,0) {$\,\,\qquad$};\node[circle,draw] at (4.2,0) {$\qquad$};
\draw [-] (x1) to (x2);
\draw [-] (x2) to (x3);
\draw [-] (x3) to (x4);
\draw [-] (x4) to (x5);
\draw [-] (x5) to (x6);
  \epic \qquad\qquad \\ 
   \bpic \draw[<->, thick] (0,0) -- (0,-1);  \node[right] at (0.3,-0.5){3d MIRROR DUAL ,\quad $\CN=4$ SUSY \quad};  \epic \label{mirrorUVakan}   \\
    \bpic  \path (-1.5,0) node {$\tilde{\CT}_{3d,UV}:$} (0.2,0) node[circle,draw](x1) {$\,1\,$} -- (1.2,0) node[circle,draw](x2) {$\,2\,$} -- (2.2,0) node(x3) {$\cdots$} -- (3.5,0) node[circle,draw](x4) {\small$\!\!kN\!\!-\!\!1\!\!$} -- (5,0) node[circle,draw](x5) {\small$\,k N\,$} --
  (3.5,1.5) node[circle,draw,red,thick](y1) {$1$} -- (4.5,1.5) node(y2) {\small{$\dots k \dots$}} -- (5.5,1.5) node[circle,draw,red,thick](y3) {$1$} -- (6.6,1.5) node[rectangle,draw](x6) {$1$} -- (6.5,0) node[circle,draw](x7) {$\!\!kN\!\!-\!\!k\!\!$} -- (7.8,0) node(x8) {$\cdots$} -- (8.8,0) node[circle,draw](x9) {\small$2 k$} -- (9.8,0) node[circle,draw](x10) {$\,k\,$};
\draw [-] (x1) to (x2);
\draw [-] (x2) to (x3);
\draw [-] (x3) to (x4);
\draw [-] (x4) to (x5);
\draw [-] (x5) to (x6);
\draw [-] (x5) to (y1);
\draw [-] (x5) to (y3);
\draw [-] (x5) to (x7);
\draw [-] (x7) to (x8);
\draw [-] (x8) to (x9);
\draw [-] (x9) to (x10);
  \epic \nn
\eea
We can understand the mirror pair using the results of \cite{Benini:2010uu} for the $3d$ mirrors of class-S $\CN=2$ theories. As we review in \ref{gaioquivers}, the quiver on top of \ref{mirrorUVakan} in $4d$ has a class-S description as $kN$ $M5$'s on a sphere with 
\begin{itemize}
\item $k+1$ minimal punctures $\circ$, 
\item one maximal puncture $\otimes$, labelled by the partition $[1^{kN}]$,
\item one puncture $\oplus$ labelled by the partition $[k^N]$. In the notation used in the Appendix it is a Young diagram with $N$ coloumns with height $k$.
\end{itemize}
As we review \ref{BTX}, the $3d$ mirror is a $\CN=4$ star-shaped quiver with $k+1$ $U(1)$-tails, and two long tails, as in the bottom of \ref{mirrorUVakan}.\footnote{Another way of deriving the mirror: instead of considering a quiver with $SU(n_i)$ gauge groups, we can gauge the $k$ $U(1)$'s to get a $U(n_i)$ quiver. $$\bpic  \path  (-1.5,0) node[circle,draw](x1) {$N$} -- (0,0) node[circle,draw](x2) {$2N$} -- (1.2,0) node(x3) {$\cdots$} -- (2.7,0) node[circle,draw](x4) {\small$\!\!\!kN\!\!-\!\!N\!\!\!$} -- (4.2,0) node[circle,draw](x5) {$\,k \, N\,$} -- (6.2,0) node[rectangle,draw](x6) {$kN\!+\!N$} ;
\draw [-] (x1) to (x2);
\draw [-] (x2) to (x3);
\draw [-] (x3) to (x4);
\draw [-] (x4) to (x5);
\draw [-] (x5) to (x6); \epic$$
We then use S-duality and Hanany-Witten rules \cite{Hanany:1996ie} to find the mirror:
$$    \bpic  \path (0.2,0) node[circle,draw](x1) {$\,1\,$} -- (1.2,0) node[circle,draw](x2) {$\,2\,$} -- (2.2,0) node(x3) {$\cdots$} -- (3.5,0) node[circle,draw](x4) {\small$\!\!kN\!\!-\!\!1\!\!$} -- (5,0) node[circle,draw](x5) {\small$\,k N\,$} -- (5,1.2) node[rectangle,draw](x6) {$k+1$} -- (6.5,0) node[circle,draw](x7) {$\!\!kN\!\!-\!\!k\!\!$} -- (7.8,0) node(x8) {$\cdots$} -- (8.8,0) node[circle,draw](x9) {\small$2 k$} -- (9.8,0) node[circle,draw](x10) {$\,k\,$};
\draw [-] (x1) to (x2);
\draw [-] (x2) to (x3);
\draw [-] (x3) to (x4);
\draw [-] (x4) to (x5);
\draw [-] (x5) to (x6);
\draw [-] (x5) to (x7);
\draw [-] (x7) to (x8);
\draw [-] (x8) to (x9);
\draw [-] (x9) to (x10);
  \epic $$
In the mirror the effect of the $\CN=4$ gauging of the $k$ $U(1)$'s is that $k$ $U(1)$'s are ungauged and a global symmetry $SU(k+1)$ appears.}

The Higgs branch global symmetry $SU(kN+N)$ of $\CT_{UV}$ is mapped to a Coulomb branch global symmetry in $\tilde{\CT}_{UV}$ which is the enhancement of the topological symmetries associated to the $kN+N-1$ nodes in the lower row of the $\tilde{\CT}_{UV}$ quiver.

We need to deform $\tilde{\CT}_{UV}$ \emph{a la} Maruyoshi-Song, on the $3d$ mirror side, the linear monopoles superpotential terms trigger a sequential confinement. We don't discuss the superpotential in detail, we just state the general result. As in \cite{Benvenuti:2017lle,Benvenuti:2017kud} all the nodes in the lower row of $\tilde{\CT}_{UV}$ confine. Starting from the leftmost $U(1)$, no new matter fields are created while dualizing the left tail:
 \be \bpic  \path  (5,0) node[circle,draw](x5) {\small$\,k N\,$} --  (3.5,1.5) node[circle,draw,red,thick](y1) {$1$} -- (4.5,1.5) node(y2) {\small{$\dots k \dots$}} -- (5.5,1.5) node[circle,draw,red,thick](y3) {$1$} -- (6.6,1.5) node[rectangle,draw](x6) {$1$}  -- (6.5,0) node[circle,draw](x7) {$\!\!kN\!\!-\!\!k\!\!$} -- (7.8,0) node(x8) {$\cdots$} -- (8.8,0) node[circle,draw](x9) {\small$2 k$} -- (9.8,0) node[circle,draw](x10) {$\,k\,$};
\draw [-] (x5) to (x6);
\draw [-] (x5) to (x7);
\draw [-] (x5) to (y1);
\draw [-] (x5) to (y3);
\draw [-] (x7) to (x8);
\draw [-] (x8) to (x9);
\draw [-] (x9) to (x10);
  \epic \ee
however, when dualyzing the right tail, at each dualization, we create hypermultiplets in the bifundamental of the groups in the upper row, $U(1)^{k} \times U(1)$. After dualizing all the nodes in the right tail, the result is precisely the complete graph $\CN=4$ with $k+1$ nodes and $N$ links:
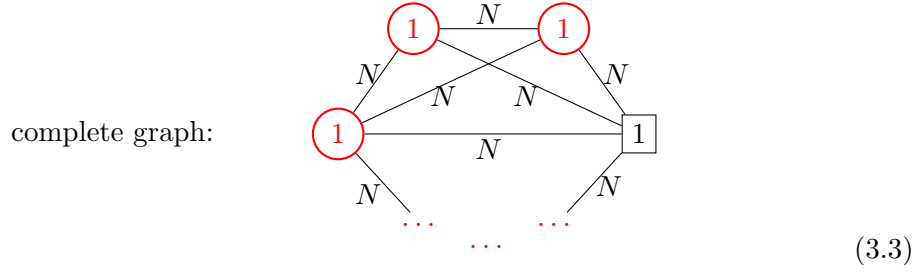
\begin{small}  \be \bpic \label{completeG}
\path (-2,0) node[circle,draw,red,thick](y1) {$1$} -- (-1,1.4) node[circle,draw,red,thick](y2) {$1$} -- (1,1.4) node[circle,draw,red,thick](y3) {$1$} -- (2,0) node[rectangle,draw](y4) {$1$} -- (-0.9,-1.2) node[red] (y5) {$\dots$} -- (0.9,-1.2) node[red] (y6) {$\dots$} -- (0,-1.5) node[red] {$\dots$};
\path (-5,0) node {complete graph:} -- (-1.6,0.8) node {$N$} -- (0,1.6) node {$N$} -- (1.7,0.8) node {$N$} -- (0,-0.2) node {$N$} -- (-1.6,-0.8) node {$N$} -- (-0.6,0.5) node {$N$}-- (0.5,0.5) node {$N$}-- (1.6,-0.7) node {$N$};
\draw [-] (y1) to (y2);\draw [-] (y1) to (y3);\draw [-] (y1) to (y4);\draw [-] (y2) to (y3);\draw [-] (y2) to (y4);\draw [-] (y4) to (y3);\draw [-] (y6) to (y4);\draw [-] (y1) to (y5);
  \epic \ee \end{small}
  
\subsection{4d analysis: global symmetries and consistent superpotential}
In $4$ dimensions we start from the quiver on the top of (\ref{mirrorUVakan}) with $\CW=\CW_{\CN=2}$ and give a maximal nilpotent vev to a gauge singlet field coupled to the Higgs Branch moment map. As expected from our $3d$ argument, we will show that the $4d$ RG flow lands on the AD theory described by $k+1$ $M5$'s on a sphere with an irregular puncture of rank-$N$ (denoted by $\star$):
\be \bpic
\draw [thick] (0,0) circle [radius=2];
\draw [->, ultra thick] (2.8,0) -- (5.2,0); 
\node at (1.2,1.7) [right] {$kN$ $M5$'s}; 
\node at (3.2,0.3) [right] {RG flow}; 
\node at (-0.1,0.8)  {$\circ$ \small{$\dots k\!+\!1 \dots$}$\circ$};
\node at (-1,-0.5) {$\oplus [k^{N}]$};
\node at (1.1,-0.5) {$\otimes [1^{kN}]$};
\node at (8.3,-0.6)  {$\star$ irreg, rank-$N$};
\node at (9.2,1.7) [right] {$k\!+\!1$ $M5$'s}; 
\draw [thick] (8,0) circle [radius=2];
\epic \ee

Giving a maximal nilpotent vev in the $SU(kN+N)$ flavor group \cite{Gadde:2013fma, Agarwal:2014rua}, we end up with a 4d $\CN=1$ theory with quiver diagram
  \be \bpic  \path (-4,0) node {$\CT_{4d, IR}:$}   (-2.5,0) node[circle,draw](x1) {$\,N\,$} -- (-0.5,0) node[circle,draw](x2) {$2N$} -- (1.5,0) node(x3) {$\cdots$} -- (3.5,0)  node[circle,draw](x5) {$k N$} -- (5.5,0) node[rectangle,draw](x6) {$\,1\,$} ;
   \path  (-2.5,0) node[circle,draw] {$\,\quad$} -- (-0.5,0) node[circle,draw] {$\,\,\quad$} -- (3.5,0)  node[circle,draw] {$\,\,\quad$};
\draw [->] (x1) to[bend left] (x2); \draw [<-] (x1) to[bend right] (x2);
\draw [->] (x2) to[bend left] (x3); \draw [<-] (x2) to[bend right] (x3);
\draw [->] (x3) to[bend left] (x5); \draw [<-] (x3) to[bend right] (x5);
\draw [->] (x5) to[bend left] (x6); \draw [<-] (x5) to[bend right] (x6);
\draw [->] (x1) to[out=-45, in=0] (-2.5,-1) to[out=180,in=225] (x1);
\draw [->] (x2) to[out=-45, in=0] (-0.5,-1) to[out=180,in=225] (x2);
\draw [->] (x5) to[out=-45, in=0] (3.5,-1) to[out=180,in=225] (x5);
\node[below right] at (-2.4,-0.6){$\phi_1$};
\node[below right] at (-0.4,-0.6){$\phi_2$};
\node[below right] at (3.6,-0.6){$\phi_k$};
\node[above right] at (-1.8,0.3){$b_1$};
\node[above right] at (0.2,0.3){$b_2$};
\node[above right] at (2.0,0.3){$b_{k-1}$};
\node[below right] at (-1.7,-0.3){$\bt_1$};
\node[below right] at (0.3,-0.3){$\bt_2$};
\node[below right] at (2.0,-0.3){$\bt_{k-1}$};
\node[above right] at (4.2,0.3){$q$};
\node[below right] at (4.3,-0.3){$\qt$};
  \epic\ee 
The superpotential reads
 \be\label{potak} \CW_{trial}= \sum_{i=1}^k tr(\phi_i (b_i \bt_i-b_{i-1}\bt_{i-1}))  + \sum_{r=0}^{kN-1} \a_r tr(\qt \phi_k^r q)\,, \ee
 where we it is understood that $b_0$ and $b_k$ are not present. We dropped terms proportional to dressed mesons $tr(\qt\phi_k^h q)$ for $h \geq kN$ because they can be written in terms of $tr(\qt\phi_k^h q)$ with $h<kN$, which are flipped to zero by the $\a_h$ singlets, so chiral ring stability, as in \cite{Benvenuti:2017lle,Benvenuti:2017kud}, implies that such terms must be dropped from $\CW$.
 
 The global symmetry is $U(1)_b^k \times U(1)_T \times U(1)_R$. The k baryonic symmetries $U(1)_b$ act with charges $\pm 1$ on the bifundamentals $b_i,\bt_i$ and on the flavor $q,\qt$. They don't mix with the R-symmetry. The symmetry $U(1)_T$ acts with charges $T[q,\qt]=-\frac{1}{2}(k+1)NT[\phi_i], T[b_i,\bt_i]=-\frac{1}{2}T[\phi_i]$ and does mix with the R-symmetry, so we have to perform A-maximization in one variable. 
 
Our convention for the trial R-charges for A-maximization \cite{Intriligator:2003jj} of the various fields is as follows \cite{Gadde:2013fma}: 
\be\label{trialr} R_{\epsilon}(\phi_i)=1+\epsilon;\; R_{\epsilon}(b_i)=\frac{1-\epsilon}{2};\; R_{\epsilon}(q)=1-(kN+N)\frac{1+\epsilon}{2};\; R_{\epsilon}(\alpha_r)=(kN+N-r)(1+\epsilon).\ee 
Using the well-known formula \cite{Anselmi:1997am}
\be a=\frac{3}{32}(3\Tr R^3-\Tr R),\ee
we find that the contribution to the trial a central charge from a hypermultiplet in the bifundamental is: 
\be a_b(\epsilon)=\frac{3}{16}\left(\frac{1+\epsilon}{2} - \frac{3}{8}(1+\epsilon)^3\right).\ee 
The contribution from a vectormultiplet is 
\be a_V(\epsilon)=\frac{3}{32}(2 + 3\epsilon^3 -\epsilon)\ee 
and that from $\alpha_r$ fields is 
\be\begin{array}{ll} a_{\alpha}(\epsilon)=\frac{3}{128}[&(-2 + 6 \epsilon^2 + 3(kN+N)^2 (1 + \epsilon)^2 + 
   3(kN+N)(-1 + \epsilon^2))\times\\
  & (kN+N-1) (kN+N + (2 + kN+N)\epsilon)].\end{array}\ee 
Finally, the contribution from $q$ and $\tilde{q}$ is 
\be a_q(\epsilon)=\frac{3}{16} kN \left(-\frac{3}{8}(1 + \epsilon)^3(kN+N)^3 + (kN+N)\frac{1 + \epsilon}{2}\right).\ee 

Combining all the contributions together, we find that the trial a central charge of our theory is 
\be\label{triala} a(\epsilon)=\frac{N^2}{3}(k^3-k)a_b(\epsilon)+\frac{k(k+1)(2k+1)N^2-6k}{6}a_V(\epsilon)+a_{\alpha}(\epsilon)+a_q(\epsilon).\ee
  
Performing A-maximization, we find that the operators $tr(\phi_i^j)$ (for $i=1,2,\ldots,k$ and $j=2,3,\ldots,N+1$) and the $\alpha_{kN-1}$ singlet violate the unitarity bound. We thus add the flipping operators $\b_{i,j}$ and remove  the $\alpha_{kN-1}$ singlet. The full superpotential becomes
 \be\label{potak1} \CW_{IR}= \sum_{i=1}^k tr(\phi_i (b_i \bt_i-b_{i-1}\bt_{i-1}))  + \sum_{r=0}^{kN-2} \a_r tr(\qt \phi_k^r q) + \!\!\sum_{\tiny{\begin{array}{c}1\leq i \leq k\\2\leq j \leq N+1\\(i,j)\neq(1,N+1) \end{array}}}\!\!\! \b_{i,j} tr(\phi_i^{\,j}) \ee
The operator $tr(\phi_1^{N+1})$ can be written in terms of $tr(\phi_1^j)$ with $j\leq N$ so it must not be flipped. A-maximization for this theory tells us all the adjoint fields $\phi_i$ have the same R-charge, independent of $k$:
\be R[\phi_i]=\frac{2}{3(N+1)}. \ee
The R-charge of other operators are
\be R[b_i,\bt_i]=1-\frac{1}{3(N+1)} \qquad R[q,\qt]=1-\frac{(k+1)N}{3(N+1)}. \ee

With these $R$-charges, the central charges $a$ and $c$ match the central charges of the $\CN=2$ AD model, computed in \cite{Xie:2012hs}.

We claim that the singlets $\b_{i,j}$ for $j \leq N$ cannot take a vev: due to quantum effects, their expectation value leads to a theory with no vacuum. This is analog to \cite{Benvenuti:2017lle,Benvenuti:2017kud} and it should be possible to prove it along the lines of \cite{Kutasov:1995np,Kutasov:1995ss}.

However, the $R=\frac{2}{3}$ operators $tr(\phi_i^{N+1})$ with $i=2,3,\ldots,k$ behave differently. The $k-1$ associated flipping fields  $\b_{i,N+1}$ can take a vev and we interpret them as (some of the) Higgs branch generators of the Argyres-Douglas theory $(A_k,A_{kN+N-1})$. The $\alpha_r$ singlets are identified with (some of the) Coulomb branch operators of the AD theory.

The following table summarizes the global symmetries of the elementary fields:
\be\label{tableANAK}
\begin{array}{c||c|c|c|c|c|c}
 & U(1)_R &  U(1)_T  & U(1)_{B_1} & \ldots & U(1)_{B_{k-1}} & U(1)_{B_k} \\ \hline
\phi_i & \frac{2}{3(N+1)} &   \frac{2}{3(N+1)} & 0 &0 & 0 & 0\\
q,\qt & 1-\frac{(k+1)N}{3(N+1)} &  -\frac{(k+1)N}{3(N+1)} & 0 &0& 0 & \pm 1\\
b_{k-1},\bt_{k-1} & 1-\frac{1}{3(N+1)} &  -\frac{1}{3(N+1)} & 0 & 0 &  \pm 1 & 0\\
\ldots &   &   &   &   &    &  \\
b_{1},\bt_{1} & 1-\frac{1}{3(N+1)} &  -\frac{1}{3(N+1)} & \pm 1 & 0 &  0 & 0\\
\a_r & \frac{2(k+1)N-2r}{3(N+1)}  &  \frac{2(k+1)N-2r}{3(N+1)}  & 0 & 0 &  0 & 0\\
\b_{i,j} & 2-  \frac{2j}{3(N+1)} &   -\frac{2j}{3(N+1)}  & 0 & 0 &  0 & 0\\
\end{array}
\ee

We normalized the $U(1)_T$ charge so that $R[\phi]=T[\phi]$. As we will see, this implies that the chiral ring elements mapped to the Coulomb branch have $R=T$, while the chiral ring elements mapped to the Higgs branch have $R=-2T$.

Notice that for $k>2$ the R-charge of the flavor $q,\qt$ can be negative. This does not violate unitarity because as we will see in section \ref{higgsbranch} the gauge-invariant operators have $R>\frac{2}{3}$: the dressed mesons $tr(\qt \phi_k^i q)$ with $R < \frac{2}{3}$ are flipped to zero, and the dressed baryons have enough insertions of $\phi$'s to make their R-charge compatible with the unitarity bound.

\subsection{Conformal manifold}\label{Cmanifold}
It is interesting to count the dimension of the conformal manifold of our $\CN=1$ Lagrangians, using the prescription of  \cite{Kol:2002zt, Benvenuti:2005wi, Green:2010da, Kol:2010ub}. Here we assume $N>1$. In the quiver
  \be \bpic  \path  (-2.5,0) node[circle,draw](x1) {$\,N\,$} -- (-0.5,0) node[circle,draw](x2) {$2N$} -- (1.5,0) node(x3) {$\cdots$} -- (3.5,0)  node[circle,draw](x5) {$k N$} -- (5.5,0) node[rectangle,draw](x6) {$\,1\,$} ;
   \path  (-2.5,0) node[circle,draw] {$\,\quad$} -- (-0.5,0) node[circle,draw] {$\,\,\quad$} -- (3.5,0)  node[circle,draw] {$\,\,\quad$};
\draw [->] (x1) to[bend left] (x2); \draw [<-] (x1) to[bend right] (x2);
\draw [->] (x2) to[bend left] (x3); \draw [<-] (x2) to[bend right] (x3);
\draw [->] (x3) to[bend left] (x5); \draw [<-] (x3) to[bend right] (x5);
\draw [->] (x5) to[bend left] (x6); \draw [<-] (x5) to[bend right] (x6);
\draw [->] (x1) to[out=-45, in=0] (-2.5,-1) to[out=180,in=225] (x1);
\draw [->] (x2) to[out=-45, in=0] (-0.5,-1) to[out=180,in=225] (x2);
\draw [->] (x5) to[out=-45, in=0] (3.5,-1) to[out=180,in=225] (x5);
\node[below right] at (-2.4,-0.6){$\phi_1$};
\node[below right] at (-0.4,-0.6){$\phi_2$};
\node[below right] at (3.6,-0.6){$\phi_k$};
\node[above right] at (-1.8,0.3){$b_1$};
\node[above right] at (0.2,0.3){$b_2$};
\node[above right] at (2.0,0.3){$b_{k-1}$};
\node[below right] at (-1.7,-0.3){$\bt_1$};
\node[below right] at (0.3,-0.3){$\bt_2$};
\node[below right] at (2.0,-0.3){$\bt_{k-1}$};
\node[above right] at (4.2,0.3){$q$};
\node[below right] at (4.3,-0.3){$\qt$};
  \epic\ee 
we need to consider the beta-functions of the $k$ gauge couplings, plus the $2k-2$ superpotential couplings associated to the interactions
\be \sum_{i=1}^k tr(\phi_i (b_i \bt_i-b_{i-1}\bt_{i-1})) \label{N=2coup}\ee
Flipping interactions are never marginal: turning on such an interaction we precisely break one gloabal symmetry, the $U(1)$ symmetry that shifts the phase of the free flipping singlet, so we do not need to consider the rest of the superpotential, which is just flipping terms.


The crucial point is the following: once the beta-functions (seen as linear functions of the anomalous dimensions of the elementary fields) for the superpotential couplings in (\ref{N=2coup}) are zero, the beta-functions for the gauge couplings automatically vanish, except for the bigger gauge group $SU(kN)$, whose beta-function fixes the scaling dimension of $q,\qt$. In other words, $k-1$ beta-functions are dependent from the others and give rise to marginal directions.
 
We conclude that the complex dimension of the conformal manifold for our $\CN=1$ quiver is $k-1$. This is precisely the complex dimension of the conformal manifold of the $\CN=2$ AD model. 

It is noteworthy that there is supersymmetry enhancement on the whole $\CN=1$ conformal manifold, it would have been logically possible that only a submanifold of the $\CN=1$ conformal manifold is actually $\CN=2$.

\subsection{Recovering the $4d$ Coulomb branch of $(A_{k},A_{kN+N-1})$}  \label{CBAnAk}
Here we identify the gauge invariant fields in $\CT^{IR}_{4D}$ that map to the generators of the Coulomb branch of the Argyres-Douglas theory $(A_k,A_{kN+N-1})$. They are pretty simple to spot:
\begin{itemize}
\item The $kN-1$ singlets $\a_r$ (for $r=0,1,\ldots,kN-2$), with $\Delta= \frac{(k+1)N-r}{N+1}$. 
\item $tr(\phi^j_i)$ (for $i=2,3,\ldots,k$, $j=N+2,N+3,\ldots,iN$), with degeneracy $k-1$ and $\Delta= \frac{j}{N+1}$.
\end{itemize}
This set of gauge invariant operators $\CO_{CB}$ satisfy $R[\CO_{CB}]=T[\CO_{CB}]$ and have no baryonic charges. In total there are 
\be kN-1+\sum_{i=2}^k (iN-N-1) =  N \frac{k(k+1)}{2} - k  \ee
such operators. In appendix \ref{Deltas} we review the scaling dimensions of the CB generators of the $(A_{k},A_{kN+N-1})$ AD model. It is easy to check that there is a one to one map between the Lagrangian operators listed above and the CB generators of $(A_{k},A_{kN+N-1})$.

\subsubsection*{Emergent superpartners}
In \cite{Benvenuti:2017kud} (see section (2.1.1)), we pointed out that each operator $\CO_{CB}$ in the Lagrangian theory which maps to the CB generators of the AD models has a superpartner $\CO'_{CB}$ under the hidden supersymmetries. Their scaling dimensions satisfy
\be \Delta[\CO_{CB}]=\Delta[\CO'_{CB}]-1.\ee
Together each pair $(\CO_{CB},\CO'_{CB})$ form an half-BPS $\CN=2$ supermultiplet that in the Dolan-Osborn notation \cite{Dolan:2002zh} are called $\mathcal{E}_{(R_{\CN=2},0,0)}$.  $\CO_{CB}$ can take a vev, while $\CO'_{CB}$ cannot take a vev.

The case $k=1$ was discussed in \cite{Benvenuti:2017kud}: the CB operators $\CO_{CB}$ are the $N-1$ $\a_r$'s, and the superpartners $\CO'_{CB}$ are precisely the $N-1$ $\b_{j}$'s. They satisfy the following relations among their superconformal R-charges:
\be R[\a_{r}] = \frac{4N-2r}{3(N+1)} = R[\b_{1, r+2}] - \frac{2}{3}\,, \qquad r=0,1,\ldots, N-1 \ee
so the relations under the emergent $\CN=2$ supersymmetry read
\be \a_r \underleftrightarrow{\quad \CN=2 \quad} \b_{1,r+2} \ee

For $k>1$ the story is a bit more complicated but it is still true. Focusing on $k=2$, the CB operators $\CO_{CB}$ are the $2N-1$ $\a_r$'s and the $N-1$ $tr(\phi^j_2)$. The superpartners $\CO'_{CB}$ turn out to be the $(N-1)$ $\b_{1,j}$'s, the $(N-1)$ $\b_{2,j}$'s, and the $N$ $tr(\bt_1 \phi_2^j b_1)$ ($j=0,1,\ldots,N-1$)
. Using the table \ref{tableANAK} it is easy to check the following relations among their superconformal R-charges:
 \be R[\a_{N+p}]=tr(\phi_2^{2N-p}) = \frac{4N-2p}{3(N+1)}   = R[\b_{i, p+2}] - \frac{2}{3}\,, \qquad i=1,2\,,\,\,p=0,1,\ldots,N-2 \ee
and
 \be R[\a_{r}] = \frac{6N-2r}{3(N+1)} = R[tr(\bt_1 \phi_2^{N-r-1} b_1)] - \frac{2}{3}\,, \qquad r=0,1,\ldots, N-1 \ee
 So we propose the following relations under the emergent $\CN=2$ supersymmetry:
\bea \{\a_{N+p},tr(\phi_2^{2N-p}) \} & \underleftrightarrow{\quad \CN=2 \quad} & \{\b_{1, p+2}, \b_{2, p+2}\} \qquad p=0,1,\ldots,N-2 \\
  \a_{r}  & \underleftrightarrow{\quad \CN=2 \quad} &  tr(\bt_1 \phi_2^{N-r-1} b_1) \qquad r=0,1,\ldots,N-1 \eea
 
We refrain from discussing the details of the cases $k>2$. The operators $\CO'_{CB}$ have a similar form to the $k=2$ case.
\subsection{The Higgs branch: dressed baryons vs $3d$ monopoles in the mirror}\label{higgsbranch}
 The $4d$ Higgs branch is equal to the $3d$ Higgs branch, which is equal to the $3d$ Coulomb branch of the $3d$ mirror \ref{completeG}, the complete graph quiver with $k+1$ nodes and $N$ links between each $U(1)$ node, with one node ungauged:
 \begin{small}  \be \bpic 
\path (-2,0) node[circle,draw,red,thick](y1) {$1$} -- (-1,1.4) node[circle,draw,red,thick](y2) {$1$} -- (1,1.4) node[circle,draw,red,thick](y3) {$1$} -- (2,0) node[rectangle,draw](y4) {$1$} -- (-0.9,-1.2) node[red] (y5) {$\dots$} -- (0.9,-1.2) node[red] (y6) {$\dots$} -- (0,-1.5) node[red] {$\dots$};
\path (-5,0) node {complete graph:} -- (-1.6,0.8) node {$N$} -- (0,1.6) node {$N$} -- (1.7,0.8) node {$N$} -- (0,-0.2) node {$N$} -- (-1.6,-0.8) node {$N$} -- (-0.6,0.5) node {$N$}-- (0.5,0.5) node {$N$}-- (1.6,-0.7) node {$N$};
\draw [-] (y1) to (y2);\draw [-] (y1) to (y3);\draw [-] (y1) to (y4);\draw [-] (y2) to (y3);\draw [-] (y2) to (y4);\draw [-] (y4) to (y3);\draw [-] (y6) to (y4);\draw [-] (y1) to (y5);
  \epic \ee \end{small}
We are going to match the generators of the $4d$ Higgs branch with the generators of the $3d$ Coulomb Branch of the complete graph Abelian quiver, which were discussed by Del Zotto and Hanany in \cite{DelZotto:2014kka} using Hilbert Series techniques \cite{Benvenuti:2006qr}.

\subsubsection*{The $k=1$ case}
We repeat the discussion of \cite{Benvenuti:2017kud}. We can make a baryon out of $N$ $q$ fields and $\binom{N}{2}$ $\phi$ fields:
  \be \CB=\varepsilon_{i_1,i_2, \ldots, i_{N}}\,\, q^{i_1} \, (\phi q)^{i_2} \, (\phi^2 q)^{i_3} \, \ldots \, (\phi^{N-1}q)^{i_{N}} \ee
  with
  \be R[\CB]= -2T[\CB] =\frac{2}{3}N\ee
a similarly defined anti-baryon $\tilde \CB$, and a meson
\be \CM=tr(\qt \phi^{N-1} q)\ee
with
\be R[\CM]=-2T[\CM]=2-\frac{4 N}{3(N+1)}+\frac{2(N-1)}{3(N+1)}=\frac{4}{3}\ee
$\CB, \tilde \CB$ and $\CM$ satisfy the chiral ring relation
 \be \CB \cdot \tilde{\CB} = \varepsilon_{i_1,i_2, \ldots, i_{N}}\,\varepsilon^{j_1,j_2, \ldots, j_{N}}\, q^{i_1} \, (\phi q)^{i_2} \, \ldots \, (\phi^{N-1}q)^{i_{N}}\, \qt_{j_1} \, (\qt \phi)_{j_2} \, \ldots \, (\qt \phi^{N-1})_{j_{N}}=\CM^{N} \,,\ee
 where we used the fact that $tr(\qt \phi^{r} q)=0$ in the chiral ring if $r<N-1$.

The chiral ring relation is precisely the defining equation of $\mathbb{C}^2/\mathbb{Z}_N$, known to be the Higgs branch of the Argyres-Douglas theory $(A_1, A_{2N-1})$.

\subsubsection*{The $k>1$ case}
We now generalize to higher $k$, we will need to use "extended baryons". We only discuss the generators since the chiral ring relations in general do not match, we expect that the chiral rings are quantum modified.

In this case there are $k$ $\Delta=1$ singlets in the $3d$ $\CN=4$ vector multiplets. The $k$ $\Delta_{3D}=1$ gauge singlets in the $\CN=4$ vector multiplets of the $3d$ mirror are mapped to the $k$ gauge invariants in our $4d$ Lagrangian: $\CM=tr(\qt \phi_k^{kN-1} q)$ and $\b_{i,N+1}$, the flipping fields for $tr(\phi_i^{N+1}), i=2,3,\ldots,k$. All these $k$ Lagrangian $4d$ operators have $R=-2T=\frac{4}{3}$.
 
 The other chiral ring generators are 'basic' monopoles, there are $2^k-1$ 'basic' monopoles with all positive topological charges (and $2^k-1$ analogous monopoles with all negative topological charges):
 \begin{itemize}
 \item $k$ monopoles $\M^{1,0,\ldots,0}, \M^{0,1,0,\ldots,0}, \ldots, \M^{0,\ldots,0,1}$. $\Delta_{3D}=k\frac{N}{2}$.
 \item $\binom{k}{2}$ monopoles with $2$ unit and $k-2$ zero topological charges. $\Delta_{3D}=2(k-1)\frac{N}{2}$.
 \item $\ldots$
 \item $\binom{k}{r}$ monopoles with $r$ unit and $k-r$ zero topological charges. $\Delta_{3D}=r(k-r+1)\frac{N}{2}$.
\item $\ldots$
\item $1$ monopole with $k$ unit topological charges, $\M^{1,1,\ldots,1}$. $\Delta_{3D}=k\frac{N}{2}$.
\end{itemize}

All these monopoles can be mapped to baryonic operators. These baryonic operator are quite complicated and we need a short-hand notation for them. The baryons with minimal R-charge, that is $R=\frac{2}{3}kN$, are constructed using either one $\varepsilon$-symbol (for either the smallest or the largest gauge groups) or two $\varepsilon$-symbols for two consecutive gauge groups:
\bea
\CB_{b_1 b_2 \ldots b_{k-1} q} &=& \varepsilon_{i_1,i_2, \ldots, i_{N}} (b_1 b_2 \ldots b_{k-1} q)^{i_1} (b_1 b_2 \ldots b_{k-1} \phi q)^{i_2} \ldots (b_1 b_2 \ldots b_{k-1} \phi^{N-1} q)^{i_N} \nn \\
\CB_{b_1 \bt_2 \ldots \bt_{k-1}\qt} &=& \varepsilon_{j_1, \ldots,j_{N}} (b_1)^{j_1}_{i_1}\ldots (b_1)^{j_N}_{i_N}  \varepsilon^{i_1,i_2, \ldots, i_{2N}}(\bt_2 \ldots \bt_{k-1} \qt)_{i_{N+1}} \ldots (\bt_2 \ldots \b_{k-1} \phi^{N-1} q)_{i_{2N}} \nn \\
  \CB_{(\b_2)^{2} \bt_3 \ldots \bt_{k-1} \qt} &=& \varepsilon_{j_1, \ldots,j_{2N}} (b_2)^{j_1}_{i_1} \ldots (b_2)^{j_{2N}}_{i_{2N}}  \varepsilon^{i_1,i_2, \ldots, i_{3N}}(\bt_3 \ldots \qt)_{i_{2N+1}} \ldots (\bt_3 \ldots \phi^{N-1} \qt)_{i_{3N}} \nn \\
 &=& \ldots \ldots \\
  \CB_{(\b_{k\!-\!1})^{k\!-\!1} \qt} &=& \varepsilon_{j_1, \ldots,j_{(k-1)N}} (b_{k-1})^{j_1}_{i_1} \ldots (b_{k-1})^{j_{(k-1)N}}_{i_{(k-1)N}}  \varepsilon^{i_1,i_2, \ldots, i_{kN}}(\qt)_{i_{(k-1)N+1}} \ldots (\phi^{N-1} \qt)_{i_{kN}} \nn \\
   \CB_{q^k} &=& \,  \varepsilon_{i_1,i_2, \ldots, i_{kN}} \, (q)^{i_{1}} \, (\phi q)^{i_2} \, (\phi^2 q)^{i_3} \, \ldots \, (\phi^{N-1} q)^{i_{kN}} \nn \eea

All these baryons satisfy  \be R[\CB]= -2T[\CB] =\frac{2}{3}kN\ee

A mapping to the monopoles with smallest possible dimension goes as follow
\be\begin{array}{|c|c|}\hline
baryon & monopole \\ \hline
\CB_{b_1 \bt_2 \ldots \bt_{k-1}\qt} & \M^{1,0,\ldots,0} \\
\CB_{(\b_2)^{2} \bt_3 \ldots \bt_{k-1} \qt} &  \M^{0,1,0\ldots,0} \\
\CB_{(\b_3)^{3} \bt_4 \ldots \bt_{k-1} \qt}  &  \M^{0,0,1,0\ldots,0} \\
\ldots & \ldots \\
  \CB_{(\b_{k\!-\!1})^{k\!-\!1} \qt} & \M^{0,\ldots,0,1,0} \\
\CB_{q^k} & \M^{0,0,\ldots,0,1} \\
 \CB_{b_1 b_2 \ldots b_{k-1} q} & \M^{1,1,\ldots,1} \\ \hline
\end{array}\ee

From the mapping of the smallest chiral ring generators it's easy to infer the map for all the other generators, for instance if $k=3$:
\be\begin{array}{|c|c||c|c|}\hline
baryon & monopole & antibaryon & antimonopole \\ \hline
\CB_{b_1 \bt_2  \qt} & \M^{1,0,0}     &  \CB_{\bt_1 b_2 q} & \M^{-1,0,0} \\
\CB_{(b_2)^{2}  \qt} &  \M^{0,1,0} & \CB_{(\bt_2)^{2} q} &  \M^{0,-1,0} \\
\CB_{q^3} & \M^{0,0,1} & \CB_{\qt^3} & \M^{0,0,-1} \\
\CB_{b_1 b_2  \qt^2} & \M^{1,1,0}     &  \CB_{\bt_1 \bt_2 q^2} & \M^{-1,-1,0} \\
\CB_{(b_2)^{2}  q^2} &  \M^{0,1,1} & \CB_{(\bt_2)^{2} \qt^2} &  \M^{0,-1,-1} \\
\CB_{b_1 \bt_2 q^2} & \M^{1,0,1} & \CB_{\qt^3} & \M^{-1,0,-1} \\
\CB_{b_1 b_2 q} & \M^{1,1,1} &  \CB_{\bt_1 \bt_2 \qt} & \M^{-1,-1,-1} \\ \hline
\end{array}\ee

This concludes our proof that the $2(2^k-1)$ algebraically independent baryons plus the $k$ operators $\CM=tr(\qt \phi_k^{kN-1} q)$ and $\b_{i,N+1}$ are mapped to the generators of the Higgs Branch of the $(A_{k},A_{kN+N-1})$ Argyres-Douglas theory.

\section{$M5$'s on a sphere with irregular and minimal puncture}\label{AkAnplusmin}
In this case we make a small change to \ref{mirrorUVakan} and start from the mirror pair:
\bea \nn 
\bpic  \path (-4.5,0) node {$\CT_{3d,UV}^{k,N}:$} (-3,0) node[rectangle,draw](x0) {$\,1\,$} -- (-1.5,0) node[circle,draw](x1) {$\!N\!+\!1\!$} -- (0.5,0) node[circle,draw](x2) {$\!2N\!+\!1\!$} -- (2,0) node(x3) {$\cdots$} -- (3.5,0) node[circle,draw](x5) {$\!kN\!+\!1\!$} -- (5.5,0) node[rectangle,draw](x6) {$\!kN\!+\!N\!+\!1\!$} ;
\node[circle,draw] at (-1.5,0) {$\qquad$};\node[circle,draw] at (0.5,0) {$\,\,\qquad$};\node[circle,draw] at (3.5,0) {$\,\,\qquad$};
\draw [-] (x1) to (x0);
\draw [-] (x1) to (x2);
\draw [-] (x2) to (x3);
\draw [-] (x3) to (x5);
\draw [-] (x5) to (x6);
  \epic \qquad\qquad \\ 
   \bpic \draw[<->, thick] (0,0) -- (0,-1);  \node[right] at (0.3,-0.5){3d MIRROR DUAL ,\quad $\CN=4$ SUSY \quad};  \epic \label{mirrorUVakan1}   \\
    \bpic  \path (-1.5,0) node {$\tilde{\CT}_{3d,UV}^{k,N}:$} (0.2,0) node[circle,draw](x1) {$\,1\,$} -- (1.2,0) node[circle,draw](x2) {$\,2\,$} -- (2.2,0) node(x3) {$\cdots$} -- (3.5,0) node[circle,draw](x4) {\small$\!\!kN\!\!$} -- (4.8,0) node[circle,draw](x5) {\small$\!\!kN\!\!+\!\!1\!\!$} --
  (3.5,1.5) node[circle,draw,red,thick](y1) {$1$} -- (4.5,1.5) node(y2) {\small{$\dots k \dots$}} -- (5.5,1.5) node[circle,draw,red,thick](y3) {$1$} -- (6.6,1.5) node[rectangle,draw](x6) {$1$} -- (6.6,0) node[circle,draw](x7) {\small$\!\!kN\!\!-\!\!k\!\!+1\!\!$} -- (8,0) node(x8) {$\cdots$} -- (9.2,0) node[circle,draw](x9) {\small$\!\!2k\!+\!1\!\!$} -- (10.5,0) node[circle,draw](x10) {\small$\!\!k\!+\!1\!\!$} -- (10.5,1.5) node[circle,draw,blue,thick](x11) {$1$};
\draw [-] (x1) to (x2);
\draw [-] (x2) to (x3);
\draw [-] (x3) to (x4);
\draw [-] (x4) to (x5);
\draw [-] (x5) to (x6);
\draw [-] (x5) to (y1);
\draw [-] (x5) to (y3);
\draw [-] (x5) to (x7);
\draw [-] (x7) to (x8);
\draw [-] (x8) to (x9);
\draw [-] (x9) to (x10);
\draw [-] (x11) to (x10);
  \epic \nn
\eea
We can understand the mirror pair exactly as in the previous section: the quiver on top of \ref{mirrorUVakan1} in $4d$ has a class-S description as $kN+1$ $M5$'s on a sphere with 
\begin{itemize}
\item $k+1$ minimal punctures, 
\item one maximal puncture,
\item one puncture labelled by the partition $[k^N,1]$. In the notation used in the Appendix it is a Young diagram with $N$ coloumns with height $k$.
\end{itemize}
It's $3d$ mirror is thus a $\CN=4$ star-shaped quiver with $k+1$ $U(1)$-tails, and two long tail as in the bottom of \ref{mirrorUVakan1}.\footnote{Another way of deriving the mirror: instead of considering a quiver with $SU(n_i)$ gauge groups, we can gauge the $k+1$ $U(1)$'s to get a $U(n_i)$ quiver and use Hanany-Witten rules:
\bea\nn\bpic  \path (-1,0) node[circle,draw](x1) {$\,1\,$} -- (0.2,0) node[circle,draw](x2) {$\,2\,$} -- (1.4,0) node(x3) {$\cdots$} -- (2.6,0) node[circle,draw](x4) {$k N$} -- (4.4,0) node[circle,draw](x5) {$\!\!\!kN\!\!+\!\!1\!\!\!$} -- (4.4,1.2) node[rectangle,draw](x6) {$k+1$} -- (6.2,0) node[circle,draw](x7) {\small$\!\!\!k\!N\!\!\!-\!\!k\!\!+\!\!1\!\!\!$} -- (7.6,0) node(x8) {$\cdots$} -- (8.8,0) node[circle,draw](x9) {\small$\!\!2k\!\!+\!\!1\!\!$} -- (10.1,0) node[circle,draw](x10) {$\!\!k\!\!+\!\!1\!\!$} -- (10.1,1.2) node[rectangle,draw](x11) {$1$};
\draw [-] (x1) to (x2);
\draw [-] (x2) to (x3);
\draw [-] (x3) to (x4);
\draw [-] (x4) to (x5);
\draw [-] (x5) to (x6);
\draw [-] (x5) to (x7);
\draw [-] (x7) to (x8);
\draw [-] (x8) to (x9);
\draw [-] (x9) to (x10);
\draw [-] (x10) to (x11);
  \epic \\ 
\bpic \draw[<->, thick] (0,0) -- (0,-1);  \node[right] at (0.3,-0.5){3D-MIRROR DUAL\qquad\qquad\qquad\qquad};  \epic     \\
\bpic  \path (-3,0) node[circle,draw](x0) {$\,\,1\,\,$} -- (-1.5,0) node[circle,draw](x1) {\small$\!\!N\!+\!1\!\!$} -- (0,0) node[circle,draw](x2) {\small $\!\!2N\!\!+\!\!1\!\!$} -- (1.2,0) node(x3) {$\cdots$} -- (2.7,0) node[circle,draw](x4) {\small$\!\!\!kN\!\!+\!\!1\!\!\!$}  -- (5.2,0) node[rectangle,draw](x6) {$kN\!\!+\!\!N\!\!+\!\!1$} ;
\draw [-] (x1) to (x0);
\draw [-] (x1) to (x2);
\draw [-] (x2) to (x3);
\draw [-] (x3) to (x4);
\draw [-] (x4) to (x5);
\draw [-] (x5) to (x6);
  \epic \qquad\qquad\qquad\nn\eea}. We now deform the theory with a maximal nilpotent vev for the $SU(kN+N+1)$ global symmetry. In the mirror quiver, the bottom row of the mirror quiver confines as in the previous section, and the low energy theory is an Abelian quiver with $k+1$ $U(1)$ gauge groups, $k$ red nodes and $1$ blue node, plus a flavor $U(1)$ node. There are $N$ bifundamentals connecting the $k$ red nodes and the flavor node among themselves, and there is one bifundamental between the blue node and the other $k+1$ nodes. For instance for $k=4$ the IR quiver is
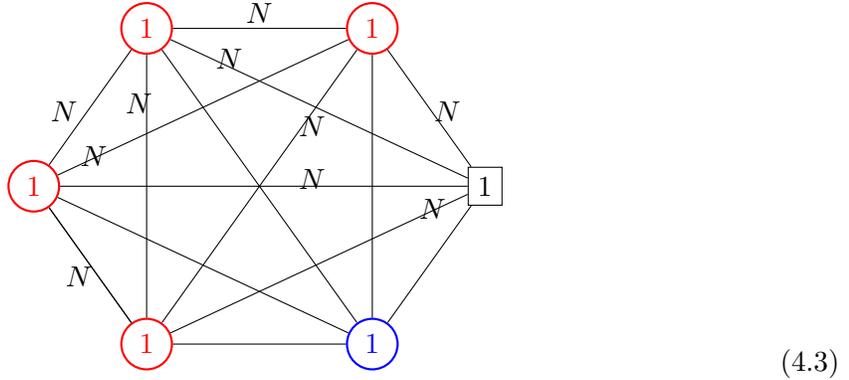
\begin{small}  \be \bpic 
\path (-3,0) node[circle,draw,red,thick](y1) {$1$} -- (-1.5,2.1) node[circle,draw,red,thick](y2) {$1$} -- (1.5,2.1) node[circle,draw,red,thick](y3) {$1$} -- (3,0) node[rectangle,draw](y4) {$1$} -- (-1.5,-2.1) node[circle,draw,red,thick] (y5) {$1$} -- (1.5,-2.1) node[circle,draw,blue,thick] (y6) {$1$};
\path (-2.6,1) node {$N$} -- (0,2.3) node {$N$} -- (2.5,1) node {$N$} -- (0.7,0.1) node {$N$} -- (-2.4,-1.2) node {$N$} -- (-1.6,1.1) node {$N$}--(-0.4,1.7) node {$N$}-- (0.7,0.8) node {$N$}--(-2.2,0.4) node {$N$} -- (2.3,-0.3) node {$N$};
\draw [-] (y1) to (y2);\draw [-] (y1) to (y3);\draw [-] (y1) to (y4);\draw [-] (y2) to (y3);\draw [-] (y2) to (y4);\draw [-] (y4) to (y3);\draw [-] (y6) to (y4);\draw [-] (y1) to (y5);\draw [-] (y1) to (y5);\draw [-] (y1) to (y6);\draw [-] (y5) to (y6);\draw [-] (y2) to (y5);\draw [-] (y2) to (y6);\draw [-] (y3) to (y5);\draw [-] (y3) to (y6);\draw [-] (y4) to (y5);
  \epic \ee \end{small}
This quiver generalizes the mirror of $(A_1,D_{2N})$ (obtained setting $k=1$), and was proposed to be the $3d$ mirror of the AD theories obtained wrapping $k+1$ $M5$'s on a sphere with an irregular and a minimal puncture.

In $4$ dimensions we start from the quiver on the top of eq. \ref{mirrorUVakan1} with $\CW=\CW_{\CN=2}$ and give a maximal nilpotent vev to a gauge singlet field coupled to the Higgs Branch moment map. As expected from our $3d$ argument, we will show that the $4d$ RG flow lands on the AD theory described by $k+1$ $M5$'s on a sphere with an irregular puncture of rank-$N$ (denoted by $\star$) and a minimal puncture:
\be \bpic
\draw [thick] (0,0) circle [radius=2];
\draw [->, ultra thick] (2.8,0) -- (5.2,0); 
\node at (1.2,1.7) [right] {$kN\!+\!1$ $M5$'s}; 
\node at (3.2,0.3) [right] {RG flow}; 
\node at (-0.1,0.8)  {$\circ$ \small{$\dots k+1 \dots$}$\circ$};
\node at (-1,-0.5) {$\oplus [k^{N},1]$};
\node at (1.1,-0.5) {$\otimes [1^{kN+1}]$};
\node at (8,1)  {$\circ$};
\node at (8.3,-0.6)  {$\star$ irreg, rank-$N$};
\node at (9.2,1.7) [right] {$k\!+\!1$ $M5$'s}; 
\draw [thick] (8,0) circle [radius=2];
\epic \ee

Uplifting the $SU(i N)$ quiver at the top of \ref{mirrorUVakan1} to $4d$ and turning on a maximal nilpotent vev for the $SU(kN+N+1)$ factor in the Higgs branch global symmetry, in the IR we are a left with the $4d$ $\CN=1$ quiver
  \be \bpic  \path (-6,0) node {$\CT^{k,N}_{4d, IR}:$} (-4.5,0) node[rectangle,draw](x0) {$1$} --  (-2.5,0) node[circle,draw](x1) {\small$\!N\!+\!1\!$} -- (-0.5,0) node[circle,draw](x2) {\small$\!\!2N\!+\!1\!\!$} -- (1.5,0) node(x3) {$\cdots$} -- (3.5,0)  node[circle,draw](x5) {\small$\!kN\!+\!1\!$} -- (5.5,0) node[rectangle,draw](x6) {$\,1\,$} ;
   \path  (-2.5,0) node[circle,draw] {$\!\qquad$} -- (-0.5,0) node[circle,draw] {$\qquad$} -- (3.5,0)  node[circle,draw] {$\,\qquad$};
\draw [->] (x1) to[bend left] (x0); \draw [<-] (x1) to[bend right] (x0);
\draw [->] (x1) to[bend left] (x2); \draw [<-] (x1) to[bend right] (x2);
\draw [->] (x2) to[bend left] (x3); \draw [<-] (x2) to[bend right] (x3);
\draw [->] (x3) to[bend left] (x5); \draw [<-] (x3) to[bend right] (x5);
\draw [->] (x5) to[bend left] (x6); \draw [<-] (x5) to[bend right] (x6);
\draw [->] (x1) to[out=-45, in=0] (-2.5,-1.2) to[out=180,in=225] (x1);
\draw [->] (x2) to[out=-45, in=0] (-0.5,-1.2) to[out=180,in=225] (x2);
\draw [->] (x5) to[out=-45, in=0] (3.5,-1.2) to[out=180,in=225] (x5);
\node[below right] at (-2.4,-0.8){$\phi_1$};
\node[below right] at (-0.4,-0.8){$\phi_2$};
\node[below right] at (3.6,-0.8){$\phi_k$};
\node[above right] at (-3.8,0.3){$b_0$};
\node[above right] at (-1.8,0.3){$b_1$};
\node[above right] at (0.2,0.3){$b_2$};
\node[above right] at (2.0,0.3){$b_{k-1}$};
\node[below right] at (-3.7,-0.3){$\bt_0$};
\node[below right] at (-1.7,-0.3){$\bt_1$};
\node[below right] at (0.3,-0.3){$\bt_2$};
\node[below right] at (2.0,-0.3){$\bt_{k-1}$};
\node[above right] at (4.2,0.3){$q$};
\node[below right] at (4.3,-0.3){$\qt$};
  \epic\ee 
Chiral ring stability implies that terms $tr(\qt\phi_k^{h}q)$ with $h>kN+1$ must be dropped from the superpotential. A-maximization resolves the mixing between a trial R-symmetry and the $U(1)_t$ global symmetry defined as in the previous section, setting also in this case
\be\label{aacc} R[\phi_i]=\frac{2}{3(N+1)} \qquad i=1,2,\ldots,k \ee 
The trial R-charge of the various fields is: 
\be\label{trial1} R_{\epsilon}(\phi_i)=1+\epsilon;\; R_{\epsilon}(b_i)=\frac{1-\epsilon}{2};\; R_{\epsilon}(q)=1-(kN+N+1)\frac{1+\epsilon}{2};\; R_{\epsilon}(\alpha_r)=(kN+N+1-r)(1+\epsilon).\ee 
As before we can write down the contribution to the trial a central charge from the fields in our theory. The contribution from hypermultiplets in the bifundamental and vectormultiplets is the same as in the previous section: 
\be a_b(\epsilon)=\frac{3}{16}\left(\frac{1+\epsilon}{2} - \frac{3}{8}(1+\epsilon)^3\right).\ee 
\be a_V(\epsilon)=\frac{3}{32}(2 + 3\epsilon^3 -\epsilon)\ee 
and that from $\alpha_r$ fields is 
\be\begin{array}{ll} a_{\alpha}(\epsilon)=\frac{3}{128} &[-2 + 6 \epsilon^2 + 3(kN+N+1)^2 (1 + \epsilon)^2 + 
   3(kN+N+1)(-1 + \epsilon^2)]\times\\
  & (kN+N) (kN+N+1 + (3 + kN+N)\epsilon).\end{array}\ee 
Finally, the contribution from $q$ and $\tilde{q}$ is 
\be a_q(\epsilon)=\frac{3}{16} (kN+1) \left(-\frac{3}{8}(1 + \epsilon)^3(kN+N+1)^3 + (kN+N+1)\frac{1 + \epsilon}{2}\right).\ee 

Combining all the contributions together, we find the trial a central charge 
\be\label{trial2} \frac{k}{3}((k^2-2)N^2+3kN+3)a_b(\epsilon)+\frac{kN(k+1)}{6}((2k+1)N+6)a_V(\epsilon)+a_{\alpha}(\epsilon)+a_q(\epsilon).\ee
By maximizing this expression, one can readily verify (\ref{aacc}) once the operator decoupling is taken into account:  
all the operators $tr(\phi_i^j)$ for $j=2,3,\ldots,N+1$ violate the unitarity bound and we need to add $(N-1)k$ $\beta$-fields to decouple them. The consistent superpotential is thus
 \be \CW = \sum_{i=1}^k tr(\phi_i (b_i \bt_i-b_{i-1}\bt_{i-1})) + \sum_{r=0}^{kN-2} \a_r tr(\qt \phi_k^r q) + \sum_{i=1}^k\sum_{j=2}^{N+1}\b_{i,j}tr(\phi_i^j) \ee
 
 The following table summarizes the non-baryonic global symmetries of the elementary fields:
\be\label{tableANAK1}
\begin{array}{c||c|c}
 & U(1)_R &  U(1)_T  \\ \hline
\phi_i & \frac{2}{3(N+1)} &   \frac{2}{3(N+1)} \\
q,\qt & 1-\frac{(k+1)N+1}{3(N+1)} &  -\frac{(k+1)N+1}{3(N+1)}\\
b_i,\bt_i & 1-\frac{1}{3(N+1)} &  -\frac{1}{3(N+1)} \\
\a_r & \frac{2(k+1)N-2r+2}{3(N+1)}  &  \frac{2(k+1)N-2r-2}{3(N+1)}  \\
\b_{i,j} & 2-  \frac{2j}{3(N+1)} &   -\frac{2j}{3(N+1)}  \\
\end{array}
\ee

 With these $R$-charges, the central charges $a$ and $c$ match those of the $\CN=2$ AD model computed in \cite{Xie:2012hs}.

The arguments of section \ref{Cmanifold} show that also for this class of theories the conformal manifold has complex dimension $k-1$.

The operators $\CO_{CB}$ are the same set of operators discussed in section \ref{CBAnAk}, plus the $k-1$ $tr(\phi_i^{iN+1})$ ($i=2,3,\ldots,k$) and one more $\a$. The generators of the Coulomb branch of the AD theory are listed in \ref{plusminim}, where it is shown that the minimal puncture adds $k$ operators to the list. These $k$ AD operators map the $k-1$ $tr(\phi_i^{iN+1})$ and $\a_0$.

In this case we refrain from discussing the emergent superpartners $\CO'_{CB}$ and the dressed baryons that map to the Higgs Branch of the AD model, the discussion should be similar to the case of $(A_k,A_{kN+N-1})$ of the previous section. 

\section{$3d$ Abelianization for the $SU(2)$ gauge theory dual to $(A_1,D_4)$}\label{3D}

When we compactify to $3d$ the $SU(2)$ theory with adjoint $\phi$ and $2$ flavors $q$ and $b$ with
\be \CW_{4d}= tr(\bt \phi b) + \a_0 tr(\qt q) + \frac{\b_2}{2} \Tr(\phi^2) \ee
a monopole superpotential is generated, proportional to $\b_2$:
\be \CW_{3d}= tr(\bt \phi b) + \a_0 tr(\qt q) + \b_2 (\frac{1}{2} \Tr(\phi^2)+ \M_{SU(2)}) \ee
We find it convenient, however, to start our study from the $3d$ theory above with the monopole term $\b_2\M$ removed. The reason is that in this case we can match completely the chiral rings, including the relations.

In four dimensions the vanishing of the gauge coupling beta-function imposes on the R-charges of elementary fields the same relation that in $3d$ is imposed by the superpotential term  $\beta_2 \M$:
\be\label{mon1}(r_q-1)+2(r_{\phi}-1)+(r_b-1)+2=0.\ee

\subsection{$SU(2)\!\!-\!\![2]$ without $\b_2\M$ vs $\CN=2$ $U(1)\!\!-\!\![3]$} \label{strangeU1}
Let us consider the theory
\be\label{flipped3d} \CW= tr(\bt \phi b) + \a_0 tr(\qt q) + \frac{\b_2}{2} \Tr(\phi^2) \ee
The global symmetry charges of the elementary fields and of the chiral ring generators are
\be\label{char22}
\begin{array}{c|ccc}
 &  U(1)_{R}^{trial} & U(1)_{B_1} & U(1)_{B_2} \\ \hline
\phi & r_{\phi} & 0 &0 \\
q,\qt & r_{q} &  \pm 1 & 0\\
b,\bt & \frac{2-r_{\phi}}{2} &  0 & \pm 1 \\ \hline\hline
\a_0,\{\M\phi\} & 2-2r_q &  0 & 0 \\ 
\M & 2-2r_q-r_{\phi} & 0 & 0 \\ \hline
\CB,\tilde{\CB}=\varepsilon \, q (\phi q) , \varepsilon \, (\qt \phi) \qt & r_{\phi}+2r_q & \pm 2 & 0 \\
\CC,\tilde{\CC}=\varepsilon \, b q , \varepsilon \, \bt \qt & r_q+\frac{2-r_{\phi}}{2} & 1 & \pm 1 \\ 
\CN,\tilde{\CN}= tr(\bt q), tr(\qt b)  & r_q+\frac{2-r_{\phi}}{2} & -1 & \pm 1 \\ 
\CM=tr(\qt \phi q)  & r_{\phi}+2r_q & 0 & 0 \\
\b_2  & 2-2r_{\phi} & 0 & 0 \\
\end{array}
\ee
where we have indicated the trial R-charge of the various fields. From Z-extremization \cite{Kapustin:2009kz,Jafferis:2010un} we find $r_q\sim0.2555$ and $r_{\phi}=0.5787$. We introduced the two baryonic symmetries $U(1)_{B_{1,2}}$. Operators like
\be \varepsilon \, q (\phi q) = \varepsilon_{a c} q^{a} \phi^{c}_{\,\,d} q^{d} \, , \qquad \varepsilon \, b q =  \varepsilon_{a c} b^{a} q^{c}\ee
are \emph{(dressed) baryons} charged under them.

 The relation $\phi^2=0$ immediately implies that dressed mesons and baryons like $\tr(\qt \phi^{n_1} q)$ and $\varepsilon (\phi^{n_2} q) (\phi^{n_3} q)$ vanish in the chiral ring if any $n_i>1$. The $\CF$-terms of $\phi$ reads
\be\label{quadrbasic} b_a\bt^c - \frac{\bt_ib^i}{2}\delta_a^c + \b_2 \phi_a^{\,\,c} = 0. \ee 
Multiplying this equation by $b_c\bt^a$ and using the $\CF$-terms $\bt\phi=\phi b=0$ we also deduce the equation $(\bt_ib^i)^2=0$.
Using the identity $\varepsilon^{ab}\delta_c^d=\varepsilon^{ad}\delta_c^b+\varepsilon^{db}\delta_c^a$ we can also show that dressed baryons of the form $\varepsilon b(\phi q)$ are zero in the chiral ring.

In conclusion, all gauge invariants in the chiral ring can be built using $q_a, \qt^a$, $(\phi q)_a, (\qt \phi)^a$, $b_a, \bt^a$, $\b_2, \a_0$, keeping track of the vanishing conditions. The list of chiral ring generators is given in the lower part of \ref{char22}. We claim that the last $8$ operators in \ref{char22} are mapped to the meson components of the abelian theory with three flavors and below we will see that the chiral ring relations are perfectly consistent with this claim.\footnote{If we turn on the monopole superpotential these become the generators of the Higgs branch of $\CN=4$ SQED with three flavors, or equivalently of $D_4$ Argyres-Douglas theory. This implies that the $U(1)_{B_1} \times U(1)_{B_2}$ global symmetry enhances to $SU(3)$.}

\subsubsection*{Quadratic relations in the chiral ring}

Using $\epsilon^{ab}\epsilon_{cd}=\delta^a_c\delta^b_d - \delta^a_d\delta^b_c$ we get four equations
\bea
\CB\tilde{\CB} &=& \CM^2 \\
\CC\tilde{\CC} &=& - \CN\tilde{\CN} \\
\CB\tilde{\CC} &=& \CN\CM \\
\CC\tilde{\CB} &=& -\tilde{\CN}\CM 
\eea
we find three more equations by contracting \ref{quadrbasic} with various operators (indicated on the lhs of the following equations)
\bea
q^a\qt_c & \rightarrow & \CN \tilde{\CN} = -\b_2\CM \label{eq5}\\ 
\varepsilon^{ad}\qt_d \qt_c & \rightarrow &  \tilde{\CC} \tilde{\CN} =- \b_2 \tilde{\CB}  \label{eq6} \\
\varepsilon_{ad}q^dq^c  & \rightarrow &  \CN \CC   = \b_2 \CB  \label{eq7}
\eea
Finally, using the relation $\varepsilon^{ab}\delta_c^d=\varepsilon^{ad}\delta_c^b+\varepsilon^{db}\delta_c^a$ we find
\bea
\tilde{\CB} \CN &=& \tilde{\CC}\CM\\
\CB \tilde{\CN} &=& -\CC\CM
\eea 
Using these equations it is possible to show that the $3\times3$ matrix 
\be\label{mesonnew} M_{SU(2)}=\left( \begin{array}{ccc}
  -\CM & -\CB &  \CC  \\
  \tilde{\CB} & \CM &  \tilde{\CN} \\
  \tilde{\CC}  & \CN & -\b_2\\
\end{array} \right) \ee 
satisfies the quadratic chiral ring identity 
\be\label{mesonrel} M_{SU(2)}^2=-\b_2 M_{SU(2)}=(\Tr M_{SU(2)})M_{SU(2)}.\ee 
We claim that this theory is IR equivalent to SQED with three flavors (denoted $Q^i$ $\tilde{Q}_i$, $i=1,2$ and $p$, $\tilde{p}$) and one singlet $\Phi$ with superpotential 
\be\label{abe13}\W=\Phi(\tilde{Q}_1Q^1+\tilde{Q}_2Q^2).\ee 
Notice that $p,\pt$ does not enter the superpotential. The mapping between the chiral ring generators is as follows: 
\be\label{char221}
\begin{array}{cc|cc}
 &  U(1)_{R} & & U(1)_R \\ \hline
\a_0,\{\M_{SU(2)}\phi\} & 2-2r_q &  \M_{U(1)}^\pm & 3-2r_{Q_i}-r_p \\ 
\M_{SU(2)} & 2-2r_q-r_{\phi} & \Phi & 2-2r_{Q_i} \\ \hline
\CB,\tilde{\CB},\CM & r_{\phi}+2r_q &  \tilde{Q}_iQ^i & 2r_{Q_i} \\
\CC,\tilde{\CC}, \CN,\tilde{\CN}  & r_q+\frac{2-r_{\phi}}{2} &\tilde{Q}_i p, \tilde{p}Q^i & r_{Q_i}+r_p \\ 
\b_2  & 2-2r_{\phi} & \pt p & 2r_p\\
\end{array}
\ee
The mapping holds when
\be r_p = 1- \rf \qquad r_{Q_i} = \frac{1}{2}\rf+r_q \ee
Using these identification the $S^3$ partition functions match as a function of $4$ variables. Notice that in the abelian theory there is a topological $U(1)$ symmetry which acts nontrivially only on the monopole operators $\M^{\pm}$. This symmetry is invisible in (\ref{flipped3d}). $\CZ$-extremization implies that the superconformal R-charges in the $U(1)$ theory are $r_{Q_i}\sim0.5451$ and $r_p\sim0.4210$

In the $U(1)$ theory the meson matrix 
\be\label{mesonU1} M_{U(1)}=\left( \begin{array}{ccc}
  \Qt_1 Q_1 & \Qt_1 Q_2  &  \Qt_1 p   \\
  \Qt_2 Q_1  & \Qt_2 Q_2  &  \Qt_2 p  \\
  \pt Q_1   & \pt Q_2  & \pt p\\
\end{array} \right) \ee 
precisely satisfies (\ref{mesonrel}). So we gave a proof that the full chiral rings, including the relations, are the same in the $SU(2)$ and in the $U(1)$ theories.

In order to get $\CN=4$ SQED with three flavors we should turn on the relevant deformation $\delta\W=\Phi\tilde{p}p$, which according to the above mapping corresponds on the $SU(2)$ side to turning on the superpotential term $\b_2\M$. In the $\CN=4$ SQED with three flavors the meson squares to zero so we expect that (\ref{mesonnew}) (modulo a redefinition of the diagonal components) satisfies the same relation once we properly take into account the superpotential term $\b_2\M$. It would be nice to study the chiral ring relations in these $3d$ $\CN=2$ theories using the techniques of \cite{Cremonesi:2013lqa, Hanany:2015via, Cremonesi:2015dja, Cremonesi:2016nbo}. It is not clear to us how to derive this result and, since the Higgs branch does not change under dimensional reduction, we expect an analogous subtlety to arise for theory (\ref{flippedir}) in four dimensions. It would be important to resolve this issue.



\subsection{$SU(2)\!\!-\!\![2]$ with $\b_2\M$ vs $\CN=4$ $U(1)\!\!-\!\![3]$} 

We will now study in detail the sphere partition function of $\CT'_{3d,IR}$ with the monopole superpotential term included: adjoint-SQCD $SU(2)$ with $2$ flavors $b,\bt$ and $q,\qt$
\be\label{kkk} \CW=tr(\bt \phi b) + \a_0 tr(\qt q)  +\b_2 tr(\phi^2) +\b_2\M \ee 
The global symmetries act on the elementary fields and $\M$ as
\be\label{charges22}
\begin{array}{c|ccccc}
 & U(1)_R & U(1)_{R-SC} &  U(1)_T  & U(1)_{B_1} & U(1)_{B_2} \\ \hline
\phi & r_{\phi}  &\frac{1}{2} &  \frac{1}{2}  & 0 &0 \\
q,\qt & r_q= \frac{2-3 r_{\phi}}{2}&\frac{1}{4} &  -\frac{3}{4} &  \pm 1 & 0\\
b,\bt & r_b= \frac{2-r_{\phi}}{2}& \frac{3}{4} &  -\frac{1}{4}  &  0 & \pm 1 \\
\b_2 & 2-2r_{\phi} & 1 &  -1  &  0 & 0 \\
\a_0 & 2-2r_q=3r_{\phi} & \frac{3}{2} &  \frac{3}{2}  &  0 & 0 \\
\M & 2r_{\phi} & 1 &  1  &  0 & 0 \\
\end{array}
\ee

The superconformal R-charges have been found performing $\CZ$-extremization for the mixing with $U(1)_T$ and are confirmed to high numerical precision. Already the fact that they are rational is a non-trivial numerical fact. The $S^3$ partition function reads
\be \CZ_{(\!(2)\!)\!-\![2]}[r_{\phi},b_1,b_2]= \s(2-2r_q)\s(2-2r_{\phi})\int_{-\infty}^{+\infty} \frac{\s(r_{\phi}\pm 2 i z)\s(r_{\phi})}{2! \s(\pm 2 i z)}\s(r_q \pm b_1 \pm i z)\s(r_b \pm b_2 \pm i z)  dz\ee
where $\s(x)=e^{l(1-x)}$ is the contribution of a chiral field and $r_q,r_b$ are given by (\ref{charges22}).

From the above analysis we expect this theory to be dual to $\CN=4$ SQED with $3$ flavors, whose $S^3$ partition function reads
\be \CZ_{(1)\!-\![3]}[r_Q,b_1,b_2]=\s(2-2r_Q)\int_{-\infty}^{+\infty} \s(r_Q \pm (b_1 + i z))\s(r_Q \pm (- b_1 + i z))\s(r_Q \pm ( b_2 + i z))  dz\ee
Notice that the FI parameter is not turned on.

We checked to very high numerical precision the equality between $\CZ_{(\!(2)\!)\!-\![2]}[\rf,b_1,b_2]$ and $\CZ_{(1)-[3]}[r_Q,b_1,b_2]$, which holds upon setting $r_Q=1-r_\phi$:
\be \CZ_{(\!(2)\!)\!-\![2]}[r,b_1,b_2] = \CZ_{(1)-[3]}[1-r,b_1,b_2] \ee

There are $8$ operators with $R=-T=1$: $ \qt \phi q$, $\b_2$ and the six 'baryons' defined previously $\CB$, $\tilde{\CB}$, $\CC$, $\tilde{\CC}$, $\CN$, $\tilde{\CN}$.
These should be identified with the Higgs branch generators, as we have explained in detail before.
The generators of the $U(1)$-theory Coulomb branch are $\M_{U(1)}^\pm$ and $\Phi$. They map to $\a_0$, the dressed monopole $\M\phi$ and to $\M$. $\M$ has $R=T=1$, $\M\phi$ and $\a_0$ have $R=T=\frac{3}{2}$. Denoting respectively with $C$ and $H$ the Cartan generators of Coulomb and Higgs $SU(2)$ symmetries of the $\CN=4$ theory, we have the identification $R=C+H$ and $T=C-H$.

\subsection{The naive dimensional reduction and its $3d$ mirror} 

As we have seen in section \ref{SC}, $\mathcal{\CT}'_{3d,UV}$ exhibits supersymmetry enhancement in the infrared and is dual to $(A_1,D_4)$ AD theory. One natural question is what happens in the IR to the naive dimensional reduction of the $4d$ UV theory, $\mathcal{T}_{3d,UV}$. The answer is totally analogous to the case of $(A_1,A_{2N-1})$ discussed in \cite{Benvenuti:2017lle,Benvenuti:2017kud}.

\be\label{MSmodel}\W_{UV}=\sum_{i=1}^4\tilde{q}_i\phi q_i+\tilde{q}_1q^{2}+\tilde{q}_2q^3+\alpha_0\tilde{q}_3q^1+ 
\alpha_1(\tilde{q}_2q^1+\tilde{q}_3q^2),\ee 
flows in the IR to
\be\label{MSir}\W_{IR}= tr(\tilde{b}\phi b) + \alpha_0 tr(\tilde{q}q)+\a_1 tr(\qt \phi q).\ee

The theory is $SU(2)$ adjoint SQCD with four flavors and superpotential (\ref{MSmodel}) and can be obtained from (\ref{flipped}) by removing the $\beta_2$ term (or equivalently by flipping $\beta_2$) and turning on $\alpha_1(\tilde{q}_2q^1+\tilde{q}_3q^2)$. On the mirror side we should accordingly replace (\ref{mirror2}) with 
\be\label{MSmirror}\W=\sum_i\varphi_i\tilde{b}_ib^i-\hat{\phi}_2(\sum_i\tilde{b}_ib^i)-\Tr(\phi_2(\sum_ib^i\tilde{b}_i))+\M^{1,0}+\M^{0,1}+\alpha_0\M^{--}+\alpha_1(\M^{-0}+\M^{0-}).\ee 
The analysis of the mirror theory proceeds exacty as in section \ref{SC} until (\ref{mirror23}), which is replaced by 
\be\W=-\frac{\gamma_2}{2}\sum_i\tilde{p}_ip_i+2\alpha_1\gamma_2+\gamma_3\det X_3+\alpha_0\gamma_3.\ee 
In this theory all the singlets $\alpha_0$, $\alpha_1$ and $\gamma_{2,3}$ become massive and the superpotential simply vanishes. We thus get theory (\ref{sqcd23}) without any singlets and zero superpotential, which is the mirror of $\CN=2$ SQED with three flavors plus three singlets $S_i$ and superpotential 
\be\label{AMSir}\W=\sum_{i=1}^3S_i\tilde{q}_iq^i.\ee 

Assuming the duality between (\ref{kkk}) and $\CN=4$ SQED of the previous subsection, we can immediately provide the abelian dual for theory (\ref{MSir}): we just need to flip the two cartan components of the meson matrix of $\CN=4$ SQED with three flavors, which according to the mapping for the chiral ring generators of the previous subsection amounts to flipping $\b_2$ and $tr(\tilde{q}\phi q)$ in $\CT'_{3d,IR}$. This operation leads precisely to the theory \ref{MSir}, the naive dimensional reduction of the Maruyoshi-Song model. This provides a clear-cut realization of the duality obstruction: the two flipping fields (\ref{AMSir}) (besides $\sum_iS_i$ which is already there in the $\CN=4$ theory) are identified in the Maruyoshi-Song model with $\Tr\phi^2$ and $\alpha_1$, which are precisely the operators which violate the unitarity bound in 4d. Their decoupling is crucial for the  supersymmetry enhancement in four dimensions however, as we clearly see here, they do not decouple in 3d obstructing supersymmetry enhancement and the duality with the $D_4$ AD theory.

\subsubsection*{$SU(2)\!\!-\!\![2]$ with flipped $\M$ vs $U(1)\!\!-\!\![3]$ with $\CW=0$} 
Starting again from the duality between (\ref{kkk}) and $\CN=4$ SQED, we can obtain an analogous duality between a $SU(2)$ theory with $2$ flavors and $U(1)$ with $3$ flavors with $\CW=0$. We need to flip the operator $\Phi$ in the Abelian side, which maps to the monopole $\M$ in SQCD. On the $SU(2)$ side the superpotential reads
\be \CW=\Tr(\bt \phi b) + \a_0 \Tr(\qt q)  +\b_2\Tr(\phi^2) + \b'_2 \M. \ee 
This fits perfectly with the analysis on the mirror side: by adapting the analysis of the mirror RG flow of section \ref{SC}, one can show that (\ref{mirror24}) becomes 
\be\W=-\frac{\gamma_2}{2}\sum_i\tilde{p}_ip_i+\beta_2(\tilde{p}_2p_2+2\tilde{p}_3p_3)+\beta'_2(\tilde{p}_2p_2+\tilde{p}_3p_3).\ee This model is a linear quiver $[1]-(1)-(1)-[1]$ where the three mesons are all flipped to zero. It is known \cite{Aharony:1997bx} that this model is the mirror of $\CN=2$ SQED with three flavors and zero superpotential.

\subsection{Flow to $SU(2)$ with $1$ flavor vs $U(1)$ with $2$ flavors}

One more check of the duality between (\ref{kkk}) and $U(1)$ with $3$ flavors $\CN=4$ is provided by giving mass to one of the two flavors in the $SU(2)$ theory. In this way we get $SU(2)$ adjoint SQCD with one flavor and rederive the duality discussed in \cite{Benvenuti:2017lle} with SQED with two flavors.

We proceed as follows: we turn on the superpotential term $tr(\bt q)$ and integrate out the two massive chirals $\bt$ and $q$. In the IR we find $SU(2)$ with adjoint $\phi$, $1$ flavor (that we call $p,\pt$) and superpotential
\be\label{defir}\CW= \a_0 tr(\pt \phi p)  +\b_2 (\Tr(\phi^2) + \M).\ee 
On the $U(1)$ side we are turning on a nilpotent mass term with rank-$1$, susy is broken to $\CN=2$ and the low energy theory is $U(1)$ with $2$ flavors ($Q,\Qt,P,\Pt$) and superpotential 
\be\label{defir2} \CW=\Phi Q\Qt +\Phi^2 P\Pt. \ee
Again, we checked numerically the equality of the associated $S^3$ partition functions
\be \CZ_{(\!(2)\!)\!-\![1]}[r_{\phi},b]= \s(2-2r_q-\rf)\s(2-2r_{\phi})\int_{-\infty}^{+\infty} \frac{\s(r_{\phi}\pm 2 i z)\s(r_{\phi})}{2! \s(\pm 2 i z)}\s(r_q \pm b \pm i z) dz\ee
\be \CZ_{(1)\!-\![2]}[r_{\Phi},b]=\s(r_{\Phi})\int_{-\infty}^{+\infty} \s(1-\frac{r_{\Phi}}{2} \pm (b + i z))\s(1-r_{\Phi} \pm (- b + i z)) dz\ee
upon setting $r_{\Phi}=2\rf$. The extremum is at $\rf=0.3481$.

\paragraph{Mapping of the chiral ring generators}
\be\label{charges222}
\begin{array}{c|ccc|c}
 & U(1)_R  &  U(1)_T  & U(1)_{B} &  \\ \hline
\phi & r_{\phi}    &  \frac{1}{2}  & 0 & \\
p,\pt & r_q= \frac{2-4 r_{\phi}}{2} &  -1 &  \pm 1 & \\ \hline \hline
tr(\pt p) & 2-4r_{\phi} &    -2  &  0 & P\Pt \\
\b_2 & 2-2r_{\phi} &   -1  &  0 & Q\Qt \\
\epsilon p \phi p & 2-3r_{\phi}   &  -\frac{3}{2}  &  +1 & Q\Pt \\
\epsilon \pt \phi \pt & 2-3r_{\phi}  &  -\frac{3}{2}  &  -1 & \Qt P \\ \hline
\a_0 & 3r_{\phi} &    \frac{3}{2}  &  0 & \M_{U(1)}^++\M_{U(1)}^- \\
\M & 2r_{\phi} &    1  &  0 & \Phi \\
\{\M\phi\} & 3r_{\phi}   &   \frac{3}{2}   &  0 & a \M_{U(1)}^- + b \M_{U(1)}^+ \\
\end{array}
\ee 
The unusual mapping for $\a_0$ is a manifestation of the fact that in the nonabelian theory the topological symmetry is emergent and will be justified shortly. We are unable to identify the precise combination of $\M^{\pm}$ to which the $SU(2)$ dressed monopole maps.

We would like to remark the following fact: the $\CF$-terms for $\Phi$ in (\ref{defir2}) gives us the chiral ring relation 
$$ Q\Qt +2\Phi P\Pt = 0$$ 
and according to the chiral ring map we have discussed, this corresponds on the $SU(2)$ side to a relation of the form 
\be\label{strangechi}\b_2 = - \M tr(\pt p).\ee 
It would be interesting to understand directly in the nonabelian theory how such a chiral ring relation arises.

\subsubsection*{Further flowing to the IR and dimensional reduction of $A_3$ AD theory}
Flipping $tr(\tilde{p}p)\leftrightarrow P\Pt$ in (\ref{defir}) we find the duality
\be\label{qcd2}SU(2)\!-\![1],\, \W=\a'\tilde{p}p+\a_0\tilde{p}\phi p+\b_2\Tr\phi^2+\b_2\M \leftrightarrow U(1)\!-\![2],\, \W=\Phi Q\Qt +\Phi^2 P\Pt+\a'P\Pt.\ee 
The second superpotential term in the r.h.s. violates chiral ring stability (due to the $\CF$-term for $\a'$), so it should be dropped from the superpotential. This fact can also be seen by mirroring twice the theory\footnote{The mirror dual is again SQED with two flavors with superpotential 
\be\W=S_1Q\Qt+S_2P\Pt+\Phi S_1+\Phi^2S_2+\a'S_2,\ee 
and integrating out massive fields we find that the superpotential simply vanishes, hence the mirror of this model is just the r.h.s. of (\ref{qcd2}) with the second superpotential term removed. See \cite{Benvenuti:2017lle} for a similar argument.}. We conclude that (\ref{qcd2}) is dual to SQED with two flipped flavors.
Notice that (\ref{qcd2}) is precisely the dimensional reduction of the Maruyoshi-Song model for $A_3$ Argyres-Douglas theory, which is dual to SQED with two flipped flavors \cite{Benvenuti:2017lle}. This is a nice further check of our duality.

In order to recover the $SU(2)$ model dual to $\CN=4$ SQED with two flavors, we should flip in (\ref{defir}) both $tr(\tilde{p}p)$ and $\a_0$. 
If we do that we find $SU(2)$ adjoint SQCD with one flavor and superpotential 
\be\label{qcd22}\W=\a' tr(\tilde{p}p) + \b_2\Tr\phi^2+\b_2\M.\ee 
By assuming that $\a_0$ is mapped to $\M^++\M^-$ as we claimed before, we are led to the conclusion that (\ref{qcd22}) is dual to SQED with two flavors and superpotential 
\be\label{abb2}\W=\Phi Q\Qt +\Phi^2 P\Pt+\b(\M_{U(1)}^++\M_{U(1)}^-)+\a'P\Pt.\ee 
At first sight this model looks rather complicated however, understanding its low energy behaviour is a simple task in the mirror dual description, since monopole operators are mapped to off-diagonal components of the meson matrix in the mirror theory.
The superpotential of the mirror dual is 
\be\W=S_1Q\Qt+S_2P\Pt+\Phi S_1+\Phi^2S_2+\b(P\Qt+Q\Pt)+\a'S_2,\ee 
which reduces, upon integrating out $\Phi$, $S_1$, $S_2$ and $\a'$ to 
\be\W=\b(P\Qt+Q\Pt).\ee 
Modulo a change of variables this model is clearly $\CN=4$ SQED with two flavors. The dual nonabelian side is the expected model discussed in \cite{Benvenuti:2017lle}
\be\W=\a'\tilde{p}p+\b_2\Tr\phi^2.\ee 
Notice that here we have dropped the monopole term because in this theory $\b_2$ is not in the chiral ring, so such a term would violate chiral ring stability. This can be seen e.g. by considering (\ref{strangechi}), which clearly reduces to $\b_2=0$ when we flip $tr(\tilde{p}p)$.

\section{Conclusions and outlook} 
In this paper we found several new examples of AD theories which admit a lagrangian UV completion. Our guiding principle in finding them is the analysis of the 3d mirror RG flow, combined with known proposals for the mirror duals of AD theories. 

It would be important to better sistematize the search for lagrangian UV completions of AD theories: our construction provides a guiding principle but crucially assumes that the lagrangian theory is a $\CN=2$ theory deformed by a nilpotent vev for a flipping field. Indeed there is a priori no reason to assume that more general constructions cannot display supersymmetry enhancement.

Once the $\CN=1$ quivers are known, it is straightforward to write down an integral expression for the superconformal index of the theories. It would be nice to analyze such indices and match with recent proposals about the superconformal index of generalized Argyres-Douglas models \cite{Buican:2017uka}, based on \cite{Buican:2015ina, Buican:2015hsa}.

As mentioned in section \ref{AkAn}, the set of theories $(A_k,A_{kN+N-1})$ are special from the S-duality point of view: they are the only ones among the class of $(G,G')$ models that display an infinite dimensional S-duality group \cite{Caorsi:2016ebt}. From our point of view, the set of theories $(A_k,A_{kN+N-1})$ is special since they admit an $\CN=1$ Lagrangian coming from a Maruyoshi-Song deformation of a $\CN=2$ quiver that, reduced to $3d$, has a mirror dual in which all non-Abelian gauge groups are balanced. The latter property is true also for the theories discussed in section \ref{AkAnplusmin}. It would be interesting to find a possible relation between these two very different points of view. It would also be interesting to study the S-duality group of the models of section \ref{AkAnplusmin}.

As for the compactification to $3d$ and the expected Abelianization as in \cite{Benvenuti:2017lle,Benvenuti:2017kud}, a new ingredient with respect to the $A_{2N-1}$ case is the generation of a monopole superpotential term in the compactification.  We only discussed the case of $SU(2)$ with $2$ flavors that Abelianizes in a non-trivial way to $U(1)$ with $3$ flavors $\CN=4$. Our analysis leaves some puzzles and it would be important to resolve them: first of all we should better understand the mirror map for quiver theories and the dynamical generation of superpotential terms in the compactification. Currently the latter is understood only for models with a single gauge group. Another issue is the analysis of the moduli space of these theories, which seems to be subject to nontrivial quantum relations. We came across this problem in section \ref{3D} studying the moduli space of the $(A_1,D_4)$ AD theory. 

An advantage of our method is that on the 3d mirror side it is relatively easy to understand which types of nilpotent vevs lead to enhancement of supersymmetry in the infrared (contrary to the four dimensional approach with a-maximization, which requires a detailed case-by-case analysis): if for instance we consider a non principal nilpotent vev for the linear quivers discussed in this paper, on the mirror side some nonabelian gauge groups will survive and even the matter content of the theory is clearly incompatible with supersymmetry enhancement. From this perspective nilpotent vevs which remove all nonabelian nodes in the 3d mirror are clearly special. 

The most natural direction for future investigations is to look for lagrangian UV completions of AD-type theories whose mirror contains nonabelian gauge groups such as Type IV theories in the notation of \cite{Xie:2012hs}. This would significantly enlarge the landscale of lagrangian UV completions of strongly coupled $\CN=2$ SCFTs. As we have already mentioned, with our procedure when a gauge group confines the adjoint chirals of neighbouring nodes in the quiver become massive and disappear from the spectrum. In principle this can be circumvented if a gauge node in the quiver is ``connected'' through bifundamental matter to two different gauge groups and both confine: in this case one will remove the adjoint chiral and the second will reintroduce it. The problem is the generation of the correct superpotential terms: in the present formulation of our procedure all superpotential terms generated along the process involve gauge singlets and these do not have the required structure to induce supersymmetry enhancement. Clearly some new ingredients are needed (presumably one should look for more general deformations besides nilpotent vevs) and at present we do not have examples of this type.

\acknowledgments{We are grateful to Matthew Buican, Sergio Cecotti and Sara Pasquetti for useful discussions. S.B. is partly supported by the INFN Research Projects GAST and ST$\&$FI and by PRIN 'Geometria delle varieta algebriche'. The research of S.G. is partly supported by the INFN Research Project ST\&FI.}

\appendix

\section{Class S theories and spectra of Argyres-Douglas Coulomb Branches}
In this appendix we review some background material about class $\mathcal{S}$ theories we need for our analysis.

\subsection{$4d$ $\CN=2$ quivers from $M5$'s on a sphere}\label{gaioquivers}
As was discussed by Gaiotto \cite{Gaiotto:2009we}, every $\CN=2$ conformal linear quiver in four dimensions with $SU(n)$ gauge groups, fundamental and bifundamental matter fields has a class $\mathcal{S}$ description: the Riemann surface is a sphere with $k+1$ minimal punctures, where $k$ is the number of gauge groups in the quiver, and two generic punctures encoding the structure of the tails at the two ends: every $\CN=2$ linear quiver consists of a ``bulk'' in which all the gauge groups have the same rank (say $N-1$) and a linear tail at both ends, with gauge groups of decreasing rank, whose structure can be described in terms of the  Young diagram with $N$ boxes associated with the puncture: the gauge group at the end of the quiver tail is $SU(l_1)$ where $l_1$ is the length of the first row of the Young diagram, the second is $SU(l')$ where $l'= l_1+l_2$ and so on. The bulk consists of $SU(N)$ gauge groups only. For example, in the case of a minimal puncture the quiver starts with a $SU(2)$ gauge group, followed by a $SU(3)$ gauge node and so on. In the case of a full puncture the quiver directly starts with a $SU(N)$ gauge node. There are bifundamental hypers between neighbouring gauge groups and the number of fundamentals at each node is fixed by the constraint $N_f=2N_c$, which ensures the vanishing of all beta functions. 

\subsection{3d mirrors of class $\mathcal{S}$ theories}\label{BTX}
The mirror duals of (the dimensional reduction of) class $\mathcal{S}$ theories of type $A_{N-1}$ with regular punctures were  worked out in \cite{Benini:2010uu}. The mirror theory is a star-shaped quiver with a central $U(N)$ gauge group with $g$ hypermultiplets in the adjoint representation (where $g$ is the genus of the Riemann surface) coupled to tails of unitary gauge groups which are in one-to-one correspondence with punctures. 

As is well known, punctures in $A_{N-1}$ class $\mathcal{S}$ theories are classified by partitions of $N$, hence they are labelled by Young diagrams with $N$ boxes. We will denote with $h_i$ the height of the i-th column and with $l_i$ the length of the i-th row. Rows and columns satisfy the constraints $l_j\leq l_i$ and $h_j\leq h_i$ for $j>i$. The structure of the linear tail in the mirror theory associated to a given puncture is dictated by the height of columns of the corresponding Young diagram as follows: we start from the central $U(N)$ node, we include a $U(k)$ gauge node with $k=\sum_{i>1}h_i$ and a hypermultiplet in the bifundamental of $U(k)\times U(N)$, then a third gauge node $U(k')$ with $k'=\sum_{i>2}h_i$ and one hyper in the bifundamental of $U(k)\times U(k')$ and so on. For example, in the case of a minimal puncture ($h_1=N-1$ and $h_2=1$) the tail consists of a single $U(1)$ node and for the full puncture ($N$ columns of height one) the tail is the so-called  $T(SU(N))$ theory: a linear quiver of $N-1$ unitary gauge groups with ranks decreasing by one unit each time as we move along the tail starting from the central node. 

\subsection{Curves and spectrum of $(A_n,A_k)$ theories}\label{Deltas}
The theory $(A_n,A_k)$ can be defined as the compactification of the $\CN=(2,0)$ theory of type $A_{k}$ (we assume without loss of generality $n\geq k$) on the sphere with one  irregular singularity of type I. The Seiberg-Witten (SW) curve and differential are 
\be\label{ciao}x^{k+1}+z^{n+1}=0;\quad\lambda_{SW}=xdz.\ee 
In the above formula the coordinate $z$ parametrizes the sphere and the puncture is located at $z=\infty$. 
Exploiting the fact that for every $\CN=2$ SCFT the SW differential has dimension one $[\lambda_{SW}]=1$, we find the constraint $[x]+[z]=1$ and imposing homogeneity of the curve we directly get 
$$[x]=\frac{n+1}{n+k+2};\quad [z]=\frac{k+1}{n+k+2}.$$ 
This can be used to determine the scaling dimension of all Coulomb branch operators, which are described as deformations of the SW curve. Using the freedom to shift $x$ and $z$ by a constant, we can remove all terms in the curve proportional to $x^n$ and $z^k$. As a consequence, in deforming the SW curve we will get terms of the form $x^{n-i}z^{k-j}u_{ij}$ ($i\leq n$ and $j\leq k$) and they all have dimension $(n+1)(k+1)/(n+k+2)$. This fact can be used to determine the dimension of $u_{ij}$, which describes a CB operator whenever $[u_{ij}]>1$. The parameters satisfying the constraint $[u_{ij}]=1$ describe mass parameters associated with (the cartan part of) the global symmetry of the theory and those with dimension strictly smaller than one are interpreted as coupling constants related to $\CN=2$ preserving relevant deformations. 

In this paper we just consider the special class $n+1=(k+1)N$, in which $[x]=N[z]=\frac{N}{N+1}$. The Coulomb branch of the theory includes operators of dimension $\frac{n}{N+1}$ with $N+1<n\leq N(k+1)$ and for every allowed value of $n$ there is at least one operator. More precisely, we have only one Coulomb branch operator for every $n$ in the range $Nk<n\leq N(k+1)$; two operators for every $n$ in the range $N(k-1)<n\leq Nk$ and so on, up to $k$ operators in the range $N+1<n\leq 2N$. Overall, we find $k(k-1+(N-1)(k+1))/2$ Coulomb branch operators. We also always have $k$ mass parameters.

\subsection{Adding a regular puncture}\label{plusminim}
We can refine the construction including one regular puncture at $z=0$. We will call the resulting theory $(I_{n+1,k+1},Y)$ where $Y$ denotes the Young diagram with $n+1$ boxes specifying the regular puncture (see also the previous subsection). The undeformed SW curve and differential are the same as in (\ref{ciao}); the difference arises in specifying the allowed deformations. In order to state the result, let us notice that the deformed SW curve can be written in the form 
$$\lambda^{n+1}=\sum_{i=2}^{n+1}\lambda^{n+1-i}\phi_i(z),$$ where $\phi_i(z)$ are meromorphic differentials of degree $i$ with poles at $z=0,\infty$. The advantage of this formulation is that the above equation is reparameterization invariant. The pole structure at infinity is the same as in (\ref{ciao}), but now we also have poles at $z=0$ so we don't have anymore the freedom to set to zero the terms proportional to $z^{k}$ by shifting $z$. The pole structure at zero is determined by the Young diagram as follows: the meromorphic differentials $\phi_i(z)$ have a pole of order $i-1$ for $2\leq i\leq l_1$, $i-2$ for $l_1<i\leq l_1+l_2$ and similarly for higher values of $i$ ($l_i$ is indeed the length of the i-th row of $Y$). In the case of minimal puncture all the differentials have a pole of order one whereas the order of the pole is $i-1$ for every $i$ in the case of maximal punctures. 

In the main body of the paper we consider models with a minimal punctures. In this case the list of Coulomb branch operators is the same as in the $(A_k,A_{N(k+1)-1})$ case, with the addition of $k$ operators whose dimension is $\frac{(1+j)N+1}{N+1}$ with $1\leq j\leq k$. We also have one extra mass parameter.

\bibliographystyle{ytphys}

\end{document}